%% file: main.tex
\documentclass[article]{aa}
\pdfoutput=1
\usepackage{lineno,hyperref}
\usepackage{natbib}
\usepackage{subfig}

\usepackage{amsmath}    % Advanced maths commands
\usepackage[dvipsnames]{xcolor} % Added by Paula
\usepackage{ae,aecompl}
\usepackage[english]{babel}

\usepackage{hyperref}
\hypersetup{colorlinks=true, citecolor=cyan, linkcolor=teal}

\usepackage{url}
\usepackage{soul}

\usepackage{graphicx}   % Including figure files
\usepackage{amsmath}    % Advanced maths commands
\usepackage{amssymb}    % Extra maths symbols
\usepackage{adjustbox}
\usepackage{threeparttable} %Footnote in tables
\usepackage{multirow} % for tables
\usepackage{makecell} % for tables
\usepackage[normalem]{ulem}
\usepackage{colortbl} % Table colors
\usepackage{txfonts}

\modulolinenumbers[5]
\usepackage[section]{placeins}

\begin{document} 

% Je kunt commando's definieeren, bv een beetje horizontale space, wat ik zelf 
% altijd handig vind in formules (anders staat alles zo dicht op elkaar):
\newcommand{\hs}{\hspace*{0.1cm}}
\newcommand{\hsm}{\hspace*{0.05cm}}
\newcommand{\Pryn}{\tt Pryngles}

%AUTHORS COMMENTS:
\newcommand{\ms}[1]{\textcolor{orange}{\textbf{MS:} #1}}
\newcommand{\ourtitle}{A general polarimetric model for transiting and non-transiting ringed exoplanets}

%-----------------------------------------------------------------------------------
%\title{The polarised side of the light curve: \\
\title{\ourtitle}
% \subtitle{.}

\author{Allard. K. Veenstra\inst{1},
        Jorge I. Zuluaga\inst{2},
        Jaime A. Alvarado-Montes\inst{3,4},
        Mario Sucerquia\inst{5,6,7},
        and Daphne M. Stam\inst{1}
          }

   \institute{Faculty of Aerospace Engineering, Technical University Delft, 
              Kluyverweg 2, 2629 HS Delft, The Netherlands\\
              \email{a.k.veenstra@tudelft.nl}
         \and
            SEAP/FACom, Instituto de F\'{\i}sica - FCEN, Universidad de Antioquia, Calle 70 No. 52-21, Medell\'in, Colombia\\
            \email{jorge.zuluaga@udea.edu.co}
            \and
            School of Mathematical and Physical Sciences, Macquarie University, Balaclava Road, North Ryde, NSW 2109, Australia
           \and
           The Macquarie University Astrophysics and Space Technologies Research Centre, Macquarie University, Balaclava Road, North Ryde, NSW 2109, Australia\\
           \email{jaime-andres.alvarado-montes@hdr.mq.edu.au}
           \and
           Instituto de F\'isica y Astronom\'ia, Facultad de Ciencias, Universidad de Valpara\'iso, Av. Gran Bretaña 1111, 5030 Casilla, Valpara\'iso, Chile
           \and
          Departamento de Ciencias, Facultad de Artes Liberales, Universidad Adolfo Ibáñez, Av.\ Padre Hurtado 750, Viña del Mar, Chile.
            \and
           N\'ucleo Milenio Formaci\'on Planetaria - NPF, Universidad de Valpara\'iso, Av. Gran Bretaña 1111, Valpara\'iso, Chile\\
           \email{mario.sucerquia@uv.cl}
             }

\date{Received June 15, 2023; accepted ??}

%-----------------------------------------------------------------------------------
% \abstract{}{}{}{}{} 
% 5 {} token are mandatory
%We focus on gas giants and investigate the influence of rings on their reflected flux and polarization curves as a function of their true anomaly. 
%The ring consists of irregularly shaped dust particles. 

\abstract
{
The detection and characterization of exorings, i.e.\ rings around exoplanets, will be the next breakthrough in exoplanetary science. It will help us to better understand the origin and evolution of planetary rings in the Solar System and beyond. Exorings are still elusive, and new and clever methods for identifying them need to be developed and tested.}
{
Polarimetry of the reflected light signals of exoplanets appears to be a promising tool to identify the tell-tale signatures of exorings. We explore the potential of polarimetry as a tool for detecting and characterizing exorings.}
{
For that purpose, we have improved the general, publicly available photometric code {\tt Pryngles} by adding the results of radiative transfer calculations with an adding-doubling algorithm that fully includes polarization; the improved package also allows for including the scattering by irregularly shaped dusty ring particles. With this improved code, we compute the total and polarized fluxes,
and the degree of polarization of model gas giant planets with or without rings,
along their orbits, for various key model parameters, such as the  
orbit inclination, ring size and orientation, ring particle albedo and optical thickness. We demonstrate the versatility of our code by predicting the total and polarized fluxes of the ``puffed-up'' planet HIP~41378~f assuming this planet
has an opaque dusty ring.}
{
Spatially unresolved dusty rings can significantly modify the flux and polarization signals of the light that is reflected by a gas giant exoplanet along its orbit.
Rings are expected to have a low polarization signal and will generally decrease the degree of polarization 
% \replaced{by blocking the polarized reflected light on the planetary atmosphere and casting shadows on it}{
along the orbit of a planet-ring system as a whole as the ring casts a shadow on the planet and/or blocks part of the light the planet reflects. During ring-plane crossings, when the thin ring is illuminated edge-on, a ringed exoplanet's flux and degree of polarization are close to those of a ring-less planet, and ring-plane crossing geometries generally appear as sharp changes in the flux and polarization curves. 
% }
Ringed planets in edge-on orbits tend to be difficult to distinguish from 
ring-less planets in reflected flux and degree of polarization. A larger
ring optical thickness, particle albedo, and/or ring width usually yields 
stronger ring characteristics in the phase functions. We also show that if HIP~41378~f is indeed surrounded by a ring, its reflected flux (compared to the star) will be of the order of $10^{-9}$, while the ring would decrease the degree of polarization in a detectable way. 
}
{
The improved version of the photometric code {\Pryn} that we present here shows that dusty rings may produce distinct polarimetric features in the light-curves of ringed giant gas planets across a wide range of orbital configurations, orientations and ring optical properties. The most diagnostic of these features occur around the ring-plane crossings as there, sharp changes in the flux and degree of polarization curves are to be expected. The short case study shows, that while HIP~41378f can not yet be directly imaged, the addition of polarimetry to future observations would aid in the characterization of the system.
}

\keywords{Planetary Systems -
          Planets and satellites: rings -
          Methods: Numerical -
          Radiative Transfer -
          Polarization }

%-----------------------------------------------------------------------------------
% If this title is too long, it will not complain but ask for another running title!
\titlerunning{A polarimetric model for transiting and non-transiting ringed exoplanets}
\authorrunning{Veenstra et al.}

%-----------------------------------------------------------------------------------
\maketitle

%-----------------------------------------------------------------------------------
% Sections:
%-----------------------------------------------------------------------------------
\input{Xmerged}

% \input{introduction.tex}
% \input{numerical.tex}
% \input{properties.tex}
% \input{results.tex}
% \input{discussion.tex}
% \input{conclusion.tex}

%-----------------------------------------------------------------------------------
% Acknowledgements:
%-----------------------------------------------------------------------------------
\begin{acknowledgements}
JIZ is supported by Minciencias Grant 1115-852-70719/70939.
\end{acknowledgements}

%-----------------------------------------------------------------------------------
% Bibliography:
%-----------------------------------------------------------------------------------
\bibliographystyle{aa}
\bibliography{biblio.bib}

%-----------------------------------------------------------------------------------
% Appendix:
%-----------------------------------------------------------------------------------
\newpage
\input{Xappendix.tex}

%-----------------------------------------------------------------------------------
\end{document}

%% file: Xmerged.tex
\section{Introduction}
\label{sec:intro}
Transit photometry has proven to be a successful technique to detect exoplanets (see, e.g. \citealt{Perryman2018, Deeg2018}). By studying the light curve of a star, such as a TESS/Kepler Interest Object (TOI or KOI), can help to determine if the star hosts exoplanets. Additionally, this approach provides valuable insights into the properties of the planets themselves. Transits mostly yield information about planetary architectures, allowing us to constrain the orbits of exoplanets and at least one of their bulk parameters, that is, their size or physical radius \citep{Seager_2003}. However, to take the biggest advantage of planetary transits, a given observer needs some `special' types of orbits (i.e. orbits near to edge-on orientations) which allow detecting the drop in stellar flux produced by the combination of observed star, extrasolar planet, and Earth.

Following the success of the transit technique in finding exoplanets, attention shifted to the geometric aspects of the signal that could reveal additional characteristics of exoplanets. These features include, among others, deviations from the spherical shape of planets (see e.g.\ \citealt{Barnes2003, Akinsanmi2020} and references therein), exomoons (see e.g.\ \citealt{Cabrera2007, Simon2007, Kipping2009, Heller2016, 2018A&A...618A.162B}), and planetary rings (see e.g.\ \citealt{Barnes2004, Zuluaga2015, deMooij2017characterizing, ohno2022framework}).

Among all the discovered exoplanets thus far, and despite that most of such planets are gas giants, the discovery of planetary rings remains elusive (e.g. \citealt{piro2020exploring}). Although the absence of rings around extrasolar planets might seem like an unsolved mystery, it could be due to the limitations of the methods and techniques we use to detect them. For instance, the fainter rings around the giant planets of our Solar System were discovered through \textit{in situ} observations by spacecraft, such as Voyager I and II for Jupiter and Neptune, and through the detection of anomalies in the light curve of a stellar occultation for Uranus \citep{charnoz2018}. However, current techniques to detect rings around exoplanets, similar to the latter method, have been unsuccessful thus far. Moreover, it is worth noting that a ringed exoplanet should project a larger area on the stellar disk than an exoplanet without rings, and so this type of exoplanet should have a larger apparent size when observed indirectly through planetary transits \citep{Zuluaga2015, zuluaga2022bright, ohno2022framework}. However, during the rest of their orbital phase, and depending on the geometry of the system, planetary rings can contribute to a noticeable increase in stellar flux \citep{Arnold2004, dyudina2005phase, Sucerquia2020, Lietzow2023}, increasing the chances of detection by an alternative route.

In light of the above, scattered/polarized light measurements have recently emerged as a crucial tool in the study of exoplanets, revealing valuable information about their properties that cannot be obtained using traditional techniques. For instance, instruments such as SPHERE/ZIMPOL intend to use light polarized from planetary surfaces to characterize cold planets \citep{Knutson2007, Schmid2018}, and other studies have proposed similar methods for the detection and characterization of directly imaged exoplanets using scattered light \citep[see, e.g.][]{stam2004using, karalidi2012looking, karalidi2013flux, Stolker2017}. All of these techniques, along with the appropriate instruments, might help us understand the nature of many extrasolar systems that show behaviors still awaiting an explanation, such as 55 Cancri e's phase curve with its time-varying occultation depth \citep{Tamburo2018, Morris2021,refId0} that defies current models of light reflection and emission \citep{Demory2022}. Or the discoveries of hot, "puffed-up" planets that appear to have extraordinarily low densities \citep{piro2020exploring}.

In 2022, \citet{zuluaga2022bright} introduced {\Pryn}\footnote{\url{https://ascl.net/2205.016}}, a novel {\tt Python} package to compute the transiting as well as the reflected light-curve of ringed exoplanets. The original version of {\tt Pryngles} used a simplified treatment of reflection and scattering and did not include polarization. In this paper, we present, test, and apply an updated version of {\Pryn} that enables the computation of total and polarized fluxes and the degree and direction of polarization for ringed exoplanets. To accomplish this we used data generated by an adding-doubling radiative transfer algorithm, which includes polarization and all orders of light scattering \citep{haan1987adding} (see Sect.~\ref{sec:light} for details).

Recently, \cite{Lietzow2023} also modeled, using a novel Monte Carlo algorithm,  the total flux and polarization of light scattered by exoplanetary rings. Although the model by Lietzow \& Wolf uses ring particles of different sizes and compositions, they assumed that all particles are homogeneous, spherical particles, which generally show much stronger angular features in their single scattering phase curves than more realistic irregularly shaped particles, as shown by e.g.\ \citet{munoz2000experimental}. Therefore, the single scattering phase curves of the particles that form the exorings in our numerical simulations are deduced from laboratory measurements of light scattered by irregularly shaped particles \citep[for a description of the experimental set-up, see][]{2012JQSRT.113..565M}. This gives a more accurate calculation of the properties of the light that is reflected or transmitted by the rings but removes the possibility for the code to easily change the composition and size of the ring particles since it is limited by laboratory measurements. In addition, all the scripts, and more importantly, a publicly available package, namely {\Pryn}, is provided along with this paper. This will allow users a quick implementation for the analysis of any candidate system with potentially ringed planets. It also allows us to explore a much wider parameter space in terms of illumination and viewing geometries and ring orientations than addressed by \cite{Lietzow2023}.

This paper is organized as follows. In Sect.~\ref{sec:num_method}, we give a brief overview of the {\Pryn} package, describe the improvements implemented for this paper, and the physics behind them. We use Sect.~\ref{sec:model} to describe the physical properties of the model planet and its ring, and Sect.~\ref{sec:results} to present and discuss the computed fluxes and polarization curves for a hypothetical Saturn-like ringed exoplanet with dusty rings, orbiting a solar-type star at a distance of at most a few AU. We apply our model and tools to study the case of the possibly ''puffed-up'' planet HIP~41378~f in Sect.~\ref{sec:case_study} and obtain very compelling results. Finally, we discuss the limitations, but also the prospects of this work in Sect.~\ref{sec:discussion}, and present a summary and draw the main conclusions of our efforts in Sect.~\ref{sec:summary_conclusion}.

%--------------------------------------------------------------------------------
%--------------------------------------------------------------------------------
%--------------------------------------------------------------------------------
\section{Numerical method}
\label{sec:num_method}

%%%%%%%%%%%%%%%%%%%%%%%%%%%%%%%%%%%%%%%%%%%%%%%%%%%%%%%%%%%%%%%%%%%%%%%%%%%%%%%%%
%%%%%%%%%%%%%%%%%%%%%%%%%%%%%%%%%%%%%%%%%%%%%%%%%%%%%%%%%%%%%%%%%%%%%%%%%%%%%%%%%
% Figure 1: illustration of the planet with ring
%%%%%%%%%%%%%%%%%%%%%%%%%%%%%%%%%%%%%%%%%%%%%%%%%%%%%%%%%%%%%%%%%%%%%%%%%%%%%%%%%
\begin{figure*}[t]
\includegraphics[width=\textwidth]{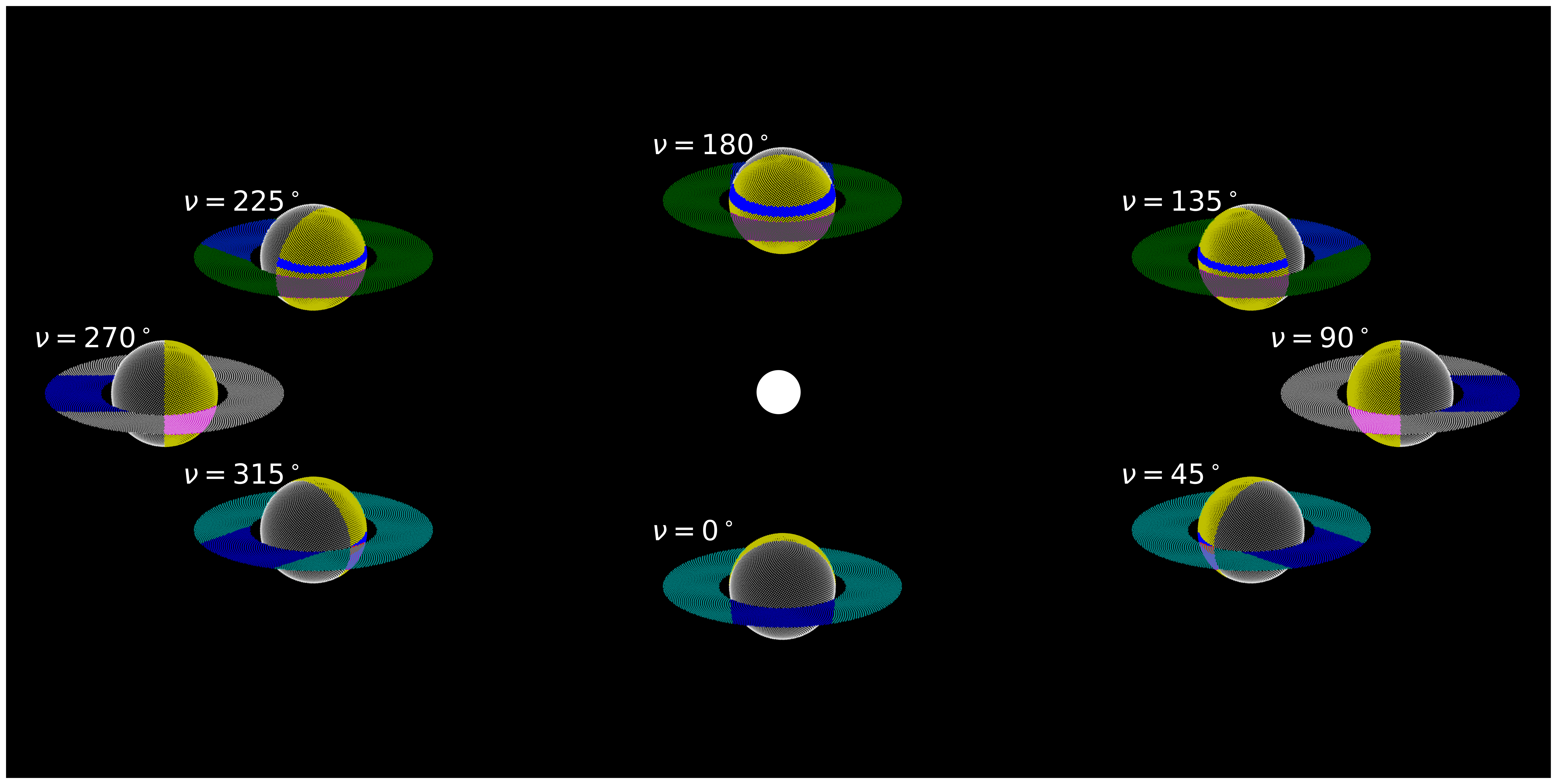}
\caption{An illustration of the various illumination and viewing 
         geometries in a planet-ring system. The star (white dot) 
         and the planetary orbit are not to scale. The circular planetary orbit 
         has an inclination angle $i$ of $60^\circ$. 
         The ring has an inclination angle $\gamma$ of $70^\circ$ 
         and an inclination longitude angle $\lambda_{\rm r}$ of $0^\circ$ 
         (this system is thus mirror--symmetric from left to right). 
         The planet's true anomaly $\nu$ is shown next to each planet's image
         and increases rotating anti-clockwise. 
         For this system, the ring-plane crossings occur at $\nu=90^\circ$ 
         and $270^\circ$. Along the part of the orbit in the upper half of
         the figure, the ring is seen in diffusely transmitted starlight (the 
         star illuminates the bottom of the ring as seen by the observer) this 
         is indicated by green-colored spangles, 
         while in the lower half, it is seen in reflected 
         starlight and indicated by cyan-colored spangles
         (see also Fig.~\ref{fig:side_view}). 
         Dark-blue spangles indicate shadows on the planet 
         and/or the ring. Yellow-colored spangles indicate planetary spangles in 
         direct starlight. Magenta-colored spangles indicate planetary spangles 
         seen through the ring and gray-colored spangles are unilluminated spangles.
         The dark-gray, small patterns on the planet 
         and the ring are a moir\'{e} pattern caused by 
         the spangles used in {\tt pryngles}.}
\label{fig:diff_positions}
\end{figure*}
%%%%%%%%%%%%%%%%%%%%%%%%%%%%%%%%%%%%%%%%%%%%%%%%%%%%%%%%%%%%%%%%%%%%%%%%%%%%%%%%%
%%%%%%%%%%%%%%%%%%%%%%%%%%%%%%%%%%%%%%%%%%%%%%%%%%%%%%%%%%%%%%%%%%%%%%%%%%%%%%%%%

Below we summarize some need-to-know characteristics of the existing {\Pryn} such as the spangles and the system geometry. A convenient illustration of the {\Pryn} approach and an overview of a ringed planet configuration is depicted in Fig.~\ref{fig:diff_positions}. We will systematically refer to this figure in the following paragraphs. After the introduction to {\Pryn} we first describe the basis of polarized light, followed by an explanation of the new way {\Pryn} models reflection/transmission of light, which is based on the adding-doubling algorithm \citep{haan1987adding}.

%--------------------------------------------------------------------------------
%--------------------------------------------------------------------------------
\subsection{The illumination and viewing geometries}
\label{sec:pryngles}

The main idea behind {\Pryn} \citep{zuluaga2022bright} is to discretize the surface (or atmosphere) 
of a planet and the surface of a ring using plane, circular area elements 
called ``spangles'', which resemble sequins or spangles found on elegant clothing (see the discrete elements in Fig.~\ref{fig:diff_positions}). The spangles are uniformly distributed over the planet and ring 
surfaces using Fibonacci spiral sampling, which prevents under- or oversampling and aliasing effects \citep{zuluaga2022bright}. Based on the illumination and viewing angles as well as shadowing and occultation by the planet or ring, {\Pryn} determines whether a spangle is 'active', in other words, if that part of the planet or ring contributes to the total amount of reflected light. If the spangle is active {\Pryn} then calculates the amount of scattered light. For a more detailed explanation of the angles involved in the calculation see Appendix \ref{appendix:1} and the original paper \citep{zuluaga2022bright}.

%%%%%%%%%%%%%%%%%%%%%%%%%%%%%%%%%%%%%%%%%%%%%%%%%%%%%%%%%%%%%%%%%%%%%%%%%%%%%%%%%
%%%%%%%%%%%%%%%%%%%%%%%%%%%%%%%%%%%%%%%%%%%%%%%%%%%%%%%%%%%%%%%%%%%%%%%%%%%%%%%%%
% Figure 2: Side view of the planet and ring
%%%%%%%%%%%%%%%%%%%%%%%%%%%%%%%%%%%%%%%%%%%%%%%%%%%%%%%%%%%%%%%%%%%%%%%%%%%%%%%%%
\begin{figure}[t]
\includegraphics[width=\hsize]{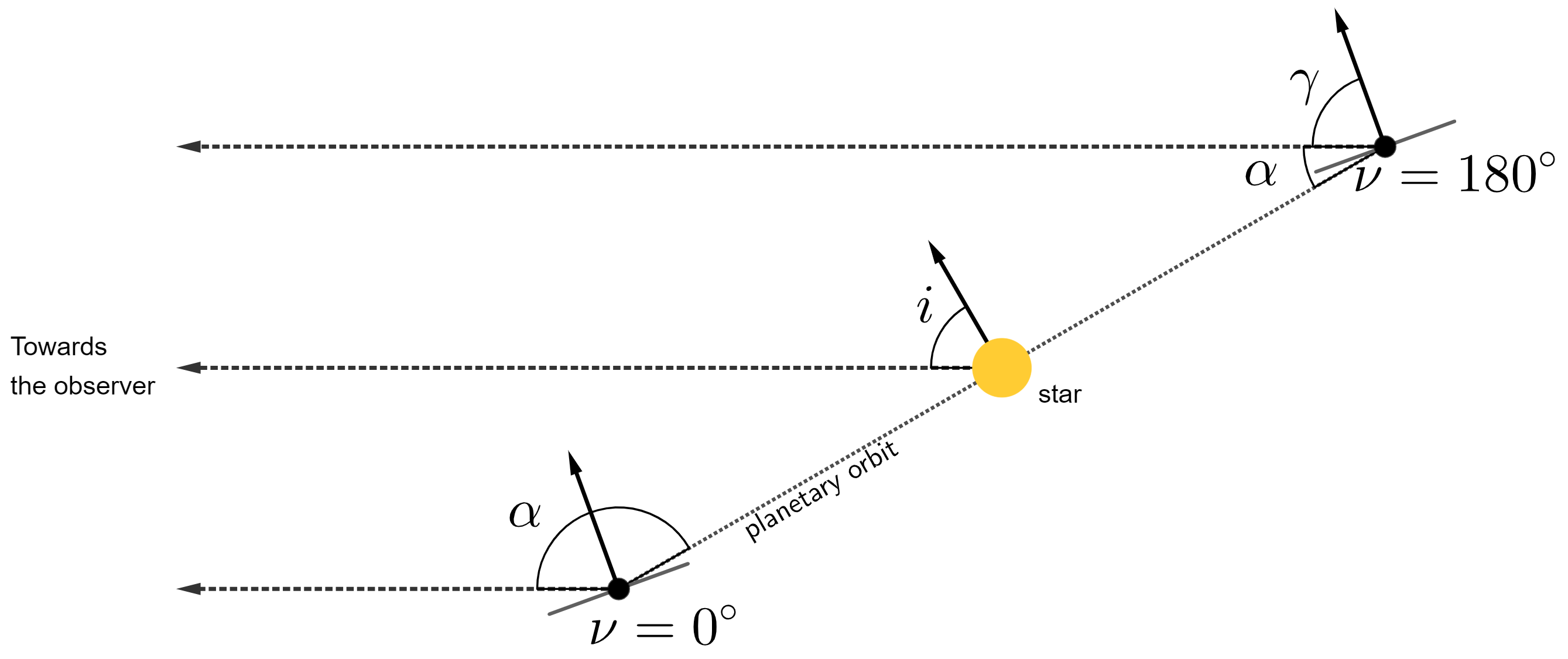}
\caption{Side-view of the planet-ring system that was also shown in 
         Fig.~\ref{fig:diff_positions} at the orbital locations where the 
         planet's true anomaly $\nu$ is 0$^\circ$ and 180$^\circ$. 
         Indicated are: the orbital 
         inclination angle $i$ and the ring inclination angle $\gamma$,
         which equal, respectively, 60$^\circ$ and 70$^\circ$. 
         Also shown is the planetary phase angle $\alpha$, which
         equals $90^\circ+i=150^\circ$ at $\nu=0^\circ$, and 
         $90^\circ-i=30^\circ$ at $\nu= 180^\circ$.}
\label{fig:side_view}
\end{figure}
%%%%%%%%%%%%%%%%%%%%%%%%%%%%%%%%%%%%%%%%%%%%%%%%%%%%%%%%%%%%%%%%%%%%%%%%%%%%%%%%%
%%%%%%%%%%%%%%%%%%%%%%%%%%%%%%%%%%%%%%%%%%%%%%%%%%%%%%%%%%%%%%%%%%%%%%%%%%%%%%%%%

The geometry of the system is defined by three angles: the orbital inclination angle $i$ ($i=90^\circ$ corresponds to an edge-on orbit), the ring inclination angle $\gamma$ ($\gamma=90^\circ$ corresponds to an edge-on ring), and the azimuthal rotation angle or roll angle $\lambda_{\rm r}$, which is only well-defined if the ring is inclined with respect to the observer (if $\lambda_{\rm r}=90^\circ$ the ring is edge-on irrespective of $\gamma$). The first two of these three angles, $i$ and $\gamma$, are clearly shown in Fig.~\ref{fig:side_view}. Examples of non-zero $\lambda_{\rm r}$ can be found on the left in Fig.~\ref{fig:geom_orbit_i20}. 

The illumination and viewing geometries of the planetary and ring spangles will vary with the location of the planet along its orbit as measured by the true anomaly $\nu$. We measure the true anomaly with respect to the closest sub-observer location in the orbit such that at $\nu=0^\circ$, the planet is between the observer and the star (if the orbital inclination $i = 90^\circ$, it would be transiting across the stellar disk); at $\nu=180^\circ$, the star is between the observer and the planet and, depending on the orbital inclination, the planet can be occulted by the star. Figure~\ref{fig:side_view} shows how $\nu$ is defined and also shows the planet's phase angle $\alpha$ which is defined as the angle between the directions to the star and the observer as measured from the center of the planet. If $\alpha=0^\circ$, the planet is fully illuminated (ignoring the possible influence of a ring and the fact that the planet would then be precisely behind its star as seen by the observer), and if $\alpha=180^\circ$, the full night-side of the planet is in view. The range of values that $\alpha$ of a planet can attain along its orbit, depends on inclination angle $i$ (see Fig.~\ref{fig:side_view}), ie. $90^\circ - i \leq \alpha \leq 90^\circ + i$. With the exception of the case when $i=0^\circ$ (a face-on orbit for which always $\alpha=90^\circ$), 
the planet's phase angle $\alpha$ changes with the true anomaly $\nu$.

Because the ring is not completely opaque, the light from the star can pass 
through it and still hit the planet (planetary spangles in the shadow of the ring will thus not
be completely dark). Light reflected by the planet can also pass through the ring 
towards the observer (the observer can 'see' the planet through the ring). It is 
also possible for light to hit the planet through the rings and then reflect toward 
the observer again through the rings. Each pass through the ring will diminish the 
flux (see Sect.~\ref{sec:results_optical_thickness}) by a factor proportional to 
the optical thickness of the ring and the illumination/viewing angle, the details 
of this will be given in Sect.~\ref{sec:integrated}. Note that light passing through the 
ring is not scattered by the ring. We thus ignore planet(ring)-shine, i.e.\ light that is first scattered by the planet (ring) and that is subsequently reflected by the ring (planet) towards the observer. Although it is an interesting second-order effect, the contribution to the total flux is mostly negligible \citep[see][]{porco2008simulations} especially for exoplanets.

Since {\Pryn} assumes a flat and geometrically zero-thickness ring (see Sect.~\ref{sec:model}), 
all the ring spangles will be dark during the so-called ring-plane crossings, i.e.\
the two locations along the orbit where the star is in the ring-plane and 
the rings do not cast a shadow on the planet. 
These crossings are always 180$^\circ$ apart. In our coordinate system, 
the true anomaly~$\nu$ at which the ring-plane crossings take place depends 
not only on the ring roll angle~$\lambda_{\rm r}$ but also on the 
ring inclination~$\gamma$ and orbital inclination~$i$ (note 
that if the ring is parallel with the planet's orbital plane, there is a perpetual
ring-plane crossing). Although at the ring-plane crossings, the rings do not cast a shadow on the planet, they can still occult some of the planet's spangles and thus still influence the total signal of the system. The location, in terms of $\nu$, where the ring-plane crossings of an arbitrary system occur can be calculated analytically (see Appendix \ref{appendix:3} for details).

%-------------------------------------------------------------------------------------
%-------------------------------------------------------------------------------------
\subsection{Physics of light scattering}
\label{sec:light}

We describe uni-directional beams of light using Stokes vectors \citep[]{HT1974,hovenier2004transfer}
\begin{equation}
    \boldsymbol{F} \equiv \begin{bmatrix} 
              F \\ Q \\ U \\ V
              \end{bmatrix}\ ,
\label{eq_stokes}
\end{equation}
with $F$ the total flux, $Q$ and $U$ the linearly polarized fluxes, 
and $V$ the circularly polarized flux, all measured in  W~m$^{-2}$ or
W m$^{-3}$ if the wavelength is included. The circularly polarized flux 
$V$ is usually very small compared to $F$, $Q$ and $U$, and will be
neglected \citep{rossi2018circular}. This does not yield significant errors 
in the computed $F$, $Q$, and $U$ \citep{stam2005errors}. 

Linearly polarized fluxes $Q$ and $U$ are defined with respect to a 
reference plane, and can be rotated from one reference plane to another using a rotation matrix ${\bf L}$ that (neglecting circular polarization) is given by \citep[see e.g.][]{hovenier2004transfer}
\begin{equation}
    \boldsymbol{L}(\beta) = 
    \begin{bmatrix}
        1 & 0 & 0 \\
        0 & \cos{2\beta} & \sin{2\beta} \\
        0 & -\sin{2\beta} & \cos{2\beta}
    \end{bmatrix},
\label{eq_rotmat}
\end{equation}
with $\beta$ the angle between the old and the new reference planes, 
measured in the anti-clockwise direction when looking to the observer. For a description of computing $\beta$ for each spangle, see Appendix ~\ref{appendix:2}.

From a Stokes vector, we obtain the polarized flux $F_{\rm pol}$ and the degree of polarization $P$, which are defined as
\begin{equation}
F_{\rm pol} \equiv \sqrt{Q^2 + U^2},
\label{eq_Fpol}
\end{equation}
and
\begin{equation}
P \equiv \frac{F_{\rm pol}}{F}.
\end{equation}
These quantities are independent of the chosen reference plane.

%-------------------------------------------------------------------------------------
%-------------------------------------------------------------------------------------
\subsection{Locally reflected and transmitted starlight}
\label{sec:locally}

To obtain the Stokes vector of light that is reflected/transmitted by the planet-ring system as a whole, we add the contributions of the individual active spangles in a similar fashion as done in Sects.~2 and~3 in \citealt{zuluaga2022bright}.
For that purpose, we need to compute first the components of the reflected or transmitted Stokes vector for every spangle. The reference plane for Stokes parameters $Q$ and $U$ is the local meridian plane of the spangle, namely the plane containing the zenith and the direction towards the observer. Note that different planetary spangles generally have different local meridian planes, while the ring spangles all have the same local meridian plane. 

The Stokes vector $\boldsymbol{F}^{\hsm \rm x}_n$ that is locally reflected by the $n-$th planetary/ring spangle is calculated according to \citep{HT1974}
\begin{equation}
\boldsymbol{F}^{\hsm \rm x}_n(\mu_{0n},\mu_n,\phi_n - \phi_{0n}) = 
   \mu_n \hs \boldsymbol{R}^{\hsm \rm x}_{n1}(\mu_{0n},\mu_n,\phi_n - \phi_{0n}) 
   \hs \mu_{0n} F_0,
\label{eq_fr}
\end{equation}
where $x$ is either `p' or `r' if the spangle belongs to the planet or ring surface respectively, $\mu_{(0)n} = \cos{\theta_{(0)n}}$ with $\theta_0$ and $\theta$ the illumination and observation angle respectively, and $\phi_n - \phi_{0n}$ is the local azimuthal difference angle (see Appendix.~\ref{appendix:1} for details). Furthermore, $\boldsymbol{R}^{\hsm \rm x}_{n1}$ is the first column of the local reflection matrix (see below), and $\pi F_0$ the flux of the incident starlight as measured on a plane perpendicular to the direction of incidence.

We assume that this starlight is uni-directional and unpolarized when integrated over the stellar disk. This assumption is based on the very small disk-integrated polarized fluxes of active and inactive FGK-stars \citep{cotton2017intrinsic} and on measurements of the Sun \citep{kemp1987optical}. Precisely, this assumption is the reason why we do not need the full reflection matrix in Eq.~\ref{eq_fr} but only its first column.

The Stokes vector for light that is locally diffusely transmitted through the $n-$th ring spangle is calculated using
\begin{equation}
\boldsymbol{F}^{\hsm \rm r}_n(\mu_{0},\mu,\phi - \phi_{0}) = 
   \mu \hs \boldsymbol{T}^{\hsm \rm r}_{n1}(\mu_{0},\mu,\phi - \phi_{0}) 
   \hs \mu_{0} F_0,
\label{eq_ft}
\end{equation}
with $\boldsymbol{T}^{\hsm \rm r}_{n1}$ the first column of the local transmission matrix. Since we assume a flat ring and parallel incident light, every ring spangle has the same illumination $\mu_0$, viewing $\mu$ and azimuthal difference angles $\phi-\phi_0$. 

The local reflection and transmission matrices $\boldsymbol{R}^{\rm p}_{n}$, $\boldsymbol{R}^{\hsm \rm r}_{n}$, and $\boldsymbol{T}^{\hsm \rm r}_{n}$ are a function of the physical properties of the medium they represent. This means that $\boldsymbol{R}^{\rm p}_{n}$ depends on the chemical composition and optical thickness of the atmosphere as well as the surface albedo. While $\boldsymbol{R}^{\hsm \rm r}_{n}$ and $\boldsymbol{T}^{\hsm \rm r}_{n}$ depend on the optical thickness of the ring and the optical properties of the particles in the ring. In this study, we assume that all planetary spangles have the same properties and all ring spangles do too, as will be described in more detail in Sect.~\ref{sec:model}.

We do not calculate reflection and transmission matrices for all individual values of $\mu$ and $\mu_0$ across the planet and the ring. Instead, we use coefficients of a Fourier expansion of $\boldsymbol{R}^{\rm p}_1$, $\boldsymbol{R}^{\hsm \rm r}_1$, and $\boldsymbol{T}^{\hsm \rm r}_1$ for various combinations of $\mu_0$ and $\mu$ (the values of Gaussian abscissae) which are pre-computed using an adding–doubling radiative transfer algorithm that fully includes polarization and all orders of scattering \citep[see][for a detailed description of the adding-doubling algorithm and the Fourier series expansion]{haan1987adding}. Finally, for values of $\mu_0$ and/or $\mu$ for which we do not have pre-calculated Fourier-coefficients, we use bi-cubic spline interpolation between pre-calculated ones, as described by \citet{rossi2018pymiedap}.

%--------------------------------------------------------------------------------
%--------------------------------------------------------------------------------
\subsection{Integrated fluxes and polarization}
\label{sec:integrated}
%{The reflected starlight of the planet-ring system}

The Stokes vector of the light that is reflected by the planet-ring 
system as a whole can be written as
\begin{equation}
\boldsymbol{F} =  \boldsymbol{F}^{\hsm \rm p} + \boldsymbol{F}^{\hsm \rm r}\,
\label{eq:total_flux}
\end{equation}
where $\boldsymbol{F}^{\hsm \rm p}$ and $\boldsymbol{F}^{\hsm \rm r}$ are the 
Stokes vectors of the planet and the ring, respectively. 

The Stokes vector $\boldsymbol{F}^{\hsm \rm p}$ of the light that is reflected
by the planet can be written as the sum of the local Stokes vectors over the
$N^{\hsm \rm p}$ active spangles on the planet, as follows
\begin{equation}
\label{eq_pl}
\boldsymbol{F}^{\hsm \rm p}(\nu) = \frac{F_0}{d^2} \hs A^{\rm p} \hs \sum^{N^{\rm p}}_{n=1} e^{-b a_n} \hs \mu_n \hs \mu_{0n} \hs \boldsymbol{L}(\beta_n) \hs \boldsymbol{R}^{\hsm \rm p}_{n1}(\mu_n,\mu_{0n},\phi_n - \phi_{0n}),
\end{equation}
with $d$ the distance to the observer,
$b$ the optical thickness of the ring (see Sect.~\ref{sec:model}),
$A^{\rm p}$ the surface area of a planetary spangle ($A^{\rm p} = 4 \pi r_{\rm p}^2 /N^{\rm p}$, with $r_{\rm p}$ the planetary radius), ${\boldsymbol L}(\beta_n)$ is the rotation matrix for the $n-$th spangle (see Eq.~\ref{eq_rotmat}), and $a_n$ a parameter that depends on whether the planetary spangle is in occultation and/or in shadow of the ring. The four possible values for $a_n$ are given in Table~\ref{tab:an}.
Matrix $\boldsymbol{L}$ in Eq.~\ref{eq_pl} rotates the local Stokes vector from 
the local meridian plane of a spangle to the reference plane of the system as a whole, 
namely the so-called ``detector plane''. This plane is fixed with respect to the 
planet and its ring, and it is equivalent to the $xz$-plane when viewed from the 
observer reference frame\footnote{This detector plane is usually different from the reference plane employed by e.g.\ \citet{stam2004using,rossi2018pymiedap} as that is the plane through the star, the planet, and the observer. That so-called ``planetary scattering plane'' is convenient for planets that are mirror-symmetric with respect to the line through the planet and the star, as then the planet's Stokes parameter $U$ equals zero. However, for a planet with a ring, $U$ will generally not equal zero with respect to the planetary scattering plane. Without this benefit, the detector plane is more directly connected to observations.}.

%%%%%%%%%%%%%%%%%%%%%%%%%%%%%%%%%%%%%%%%%%%%%%%%%%%%%%%%%%%%%%%%%%%%%%%%%%%%%%%%
\begin{table}[t]
\caption{Parameter $a_n$ (see Eq.~\ref{eq_pl}) for each of the four possible situations, with $\theta_{0}^{\hsm \rm r}$ and $\theta^{\hsm \rm r}$ the illumination and viewing angles for the ring spangles, respectively.}
\begin{center}
\begin{tabular}{|l|l|}
\hline
Situation & $a_n$ \\ \hline
No occultation or shadow & 0.0 \\
Occultation & $1/\cos{\theta^{\rm r}}$  \\
Shadow & $1/\cos{\theta_0^{\rm r}}$\\
Occultation and shadow & $1/\cos{\theta^{\rm r}} + 1/\cos{\theta_0^{\rm r}}$\\
\hline
\end{tabular}
\end{center}
\label{tab:an}
\end{table}
%%%%%%%%%%%%%%%%%%%%%%%%%%%%%%%%%%%%%%%%%%%%%%%%%%%%%%%%%%%%%%%%%%%%%%%%%%%%%%%%

The Stokes vector of the ring, $\boldsymbol{F}^{\hsm \rm r}$ (see Eq.~\ref{eq:total_flux}), is given by 
\begin{equation}
\boldsymbol{F}^{\hsm \rm r}(\nu)= \frac{F_0}{d^2} \hs A^{\rm r} \hs \sum^{N^{\rm r}}_{n=1} 
\mu \hs \mu_0 \hs \boldsymbol{L}(\beta) \hs \boldsymbol{R}^{\hsm \rm r}_{1}(\mu,\mu_0,\phi-\phi_0),
\end{equation}
for the part of the orbit where the ring reflects the incident starlight, with $N^{\rm r}$ the number of active ring spangles, or as
\begin{equation}
\boldsymbol{F}^{\hsm \rm r}(\nu) = \frac{F_0}{d^2} \hs A^{\rm r} \hs \sum^{N^{\rm r}}_{n=1} \mu \hs \mu_0 \hs \boldsymbol{L}(\beta) \hs \boldsymbol{T}^{\hsm \rm r}_1(\mu,\mu_0,\phi - \phi_0),
\end{equation}
for the part of the orbit where the ring is seen in transmitted light. All ring spangles have the same surface area, $A^{\rm r}$, which is not necessarily the same as the surface area of the planetary spangles. For a circular ring with an outer radius $r_{\rm out}$ and an inner radius $r_{\rm in}$, the spangle area is defined as $A^{\rm r}=\pi (r_{\rm out}^2 - r_{\rm in}^2)/N^{\rm r}$. 

This paper can be seen as an addition to the previous version of {\Pryn} for which the discretization and geometry have already been validated. The adding-doubling algorithm that is used to accurately calculate scattered light is the same code as found in \cite{rossi2018pymiedap} which has also been validated. To assess if the merger of the two methods has been done correctly, a comparison with a Lambertian reflecting planet has been done which can be found in Appendix \ref{appendix:4}.

In the results presented in this paper, the total and polarized fluxes, $F$ and $F_{\rm pol}$, computed by summing up the contribution of reflected/transmitted star-light from all spangles, are normalized such that at a phase angle $\alpha$ of $0^\circ$, total flux $F$ equals the ringless planet's geometric albedo $A_{\rm G}$\footnote{The geometric albedo $A_{\rm G}$ of a planet is the ratio of the total flux obtained from the planet at $\alpha=0^\circ$ and the total flux obtained from a white, Lambertian (i.e.\ isotropically) reflecting disk extending the same solid angle in the sky. For a white planet with a Lambertian reflecting surface, $A_{\rm G} = 2/3$.}.
The fluxes we compute and report in the following sections are thus dimensionless but actual fluxes received from a given planetary system, measured in W m$^{-2}$ or Jansky, can be obtained by multiplying them with the following factor $F_{\rm norm}$
\begin{equation}
    F_{\rm norm}\equiv\frac{F_0 \hs r_{\rm p}^2}{4\hs d^2},
\end{equation}
with $\pi F_0$ the flux of the starlight that is incident on the planet, 
$r_{\rm p}$ the radius of the planet, and $d$ the distance between the system and the observer. This normalization is done in order to shrink the parameter space while still clearly demonstrating the impact that rings have on the flux and polarization curves without loss of information as the results could easily be scaled to a "real system" that has the same orbital inclination, ring geometry, and physical ring/planet properties. 

Furthermore, this approach can be helpful when trying to solve the inverse problem because the distance between the planet and the star, the eccentricity of the orbit, and the stellar luminosity could be solved for, without redoing the more costly radiative transfer calculations.

%--------------------------------------------------------------------------------
%--------------------------------------------------------------------------------
%--------------------------------------------------------------------------------
\section{Properties of the model planet and its ring}
\label{sec:model}

We use a simple model planet in a circular orbit to focus on the influence of a
ring on the reflected flux and polarization signals. 
The planet is spherical and has a gaseous atmosphere that is 
bounded below by a Lambertian reflecting surface with an albedo of 0.5, 
to mimic a deep cloud layer. The gas molecules in the atmosphere 
are anisotropic Rayleigh scatterers with a depolarization factor of 
0.02, a typical value for H$_2$ \citep{HT1974}. 
In future studies, clouds and/or absorbing molecules like CH$_4$ 
could be added to simulate more complex planetary atmospheres, such as
those of the gas giants in our Solar System. 
Examples of the effects of adding clouds or absorbing methane gas 
on the total flux and polarization signals of planets can be found in 
for example \citet{stam2004using, karalidi2012looking, karalidi2013flux, rossi2018pymiedap}. 
Similarly, the effect of planetary oblateness on the flux and polarization 
could be added \citep[see, e.g.][for results of total fluxes reflected by flattened, Saturn-like planets]{dyudina2005phase} 
and the effects of clouds combined with oblateness on the flux and degree of 
polarization of emitted thermal fluxes has been studied by 
\citet{Stolker2017}. 

The ring we use is flat, circular, and horizontally homogeneous.
While geometrically the ring is infinitely thin, it is composed of irregularly shaped particles. 
The total flux and polarization of light that interacts with the 
ring depends on the illumination and viewing geometries, 
on the ring optical thickness $b$, and the optical 
properties of the ring particles like their single scattering 
albedo~$\varpi$ and their scattering matrix \citep{mishchenko2009electromagnetic}. 
The optical thickness of our model ring will vary between 0.01 and 4.0. 
This is approximately the range of optical thicknesses at visible wavelengths found across  
Saturn's horizontally inhomogeneous rings \citep{planetary_science_Pater}. 

Both the single scattering albedo and scattering matrix depend on the size, shape, and 
composition of the particles, and, for non-spherical particles, on their orientation. 
Our model ring particles are irregularly shaped and randomly oriented. Using irregularly 
shaped particles is important to avoid sharp angular features 
such as rainbows or glories that arise when using spherical 
particles \citep{goloub2000cloud, nousiainen2012comparison}. 

We only use the data of one type of particle in our calculations to keep the number of varied 
parameters down. This is justified by the fact that the change in behavior when going from a 
spherical particle to an irregular particle is generally much larger than between irregular 
particles of different composition, provided the size is the same for all three \citep{nousiainen2012comparison}. 
This holds especially true when computing the degree of polarization of the reflected light. 

The single scattering properties of our ring particles are based 
on laboratory measurements of light that is 
scattered by olivine particles with sizes that are described by
a log-normal distribution \citep{munoz2000experimental}. 
The scattering matrix of these particles was calculated by \citet{moreno2006scattering}
using the Discrete Dipole Approximation (DDA) method 
\citep{draine1994discrete, draine2004user} to fit the measurements. 
The light scattering measurements are available at 442~and 633~nm. 
We use the 633~nm data, keeping in mind the wavelength region and
capabilities of JWST \citep{rieke2005overview,jakobsen2022nearinfrared}. 
We select the particles that \citet{moreno2006scattering} call the `shape~5' particles, with an 
average projected surface area of 4.2~$\mu$m$^2$ and equivalent radii of up to 
1~$\mu$m to use in our calculations. For use in our adding-doubling radiative transfer algorithm, we expand the 
scattering matrix elements of the ring particles 
into generalized spherical functions \citep{rooij1984expansion}. 

Due to the chosen particle size, the modeled ring will be more akin to the E-ring of
Saturn, which has particle sizes between 0.2 and 10~$\mu$m \citep{Ye2016In-situ}. 
Using larger, macroscopic particles would mean that their mutual shadowing 
would have to be taken into account. This is not needed to illustrate the 
basic effects of a ring on the reflected flux and degree of polarization
of a planet and would also require a different radiative transfer approach. 

\begin{centering}
\begin{figure}
\includegraphics[width=\hsize]{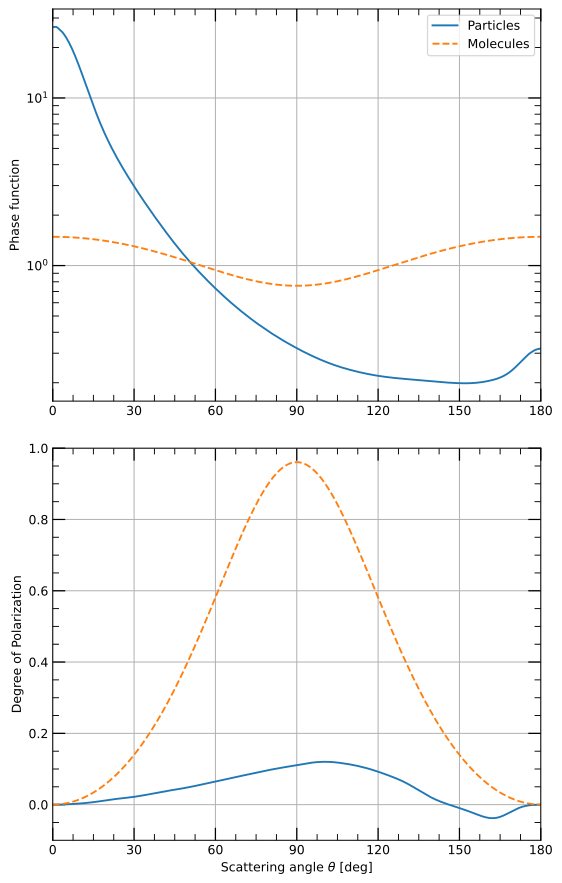}
\caption{The phase function or flux (top) and degree of polarization 
         (bottom) of incident unpolarized light that has been singly 
         scattered by gas molecules (orange, dashed) 
         and the irregularly shaped ring particles 
         \citep{moreno2006scattering} 
         (blue, solid) as functions of the single scattering angle $\Theta$ ($\Theta=0^\circ$ for forward scattered light). 
         The fluxes have been normalized such their average
         over all scattering directions equals one \citep{HT1974}. 
         Positive (negative) polarization indicates a direction of 
         polarization perpendicular (parallel) to the plane through the 
         incident and scattered light beams.}
\label{fig:single_scattering}
\end{figure}
\end{centering}

The flux $F$ and degree of polarization $P$ of the light that is singly scattered
by the ring particles and the gaseous molecules that form the planetary
atmosphere is shown in Fig.~\ref{fig:single_scattering}. 
Note that in the measurements by \citet{munoz2000experimental}, the phase functions have 
not been absolutely calibrated since the number of particles in the 
aerosol beam in the laboratory experiment is not known.
The singly scattered fluxes, which are also called the phase functions, 
have been normalized such that their average overall 
scattering directions equals one \citep{HT1974}.

%%%%%%%%%%%%%%%%%%%%%%%%%%%%%%%%%%%%%%%%%%%%%%%%%%%%%%%
%%%%%%%%%%%%%%%%%%%%%%%%%%%%%%%%%%%%%%%%%%%%%%%%%%%%%%%

%--------------------------------------------------------------------------------
%--------------------------------------------------------------------------------
%--------------------------------------------------------------------------------
\section{Results}
\label{sec:results}
Before presenting and discussing the reflected light signals of 
ringed planets, we discuss those of a planet without a ring in a 
circular orbit around its star in Sect.~\ref{sec:ring_less},
in order to better highlight the changes to the light curves that can be expected 
when a ring is present.

To explore the influence of a ring on the reflected light of a 
planet-ring system, we will now define a "standard system" that will be used to 
vary different properties and observe their respective influence on the light curves.
The standard system is made up of the same model planet 
as in the ring-less case (again on a circular orbit) but has a ring 
with $r_{\rm in}=1.20$ and $r_{\rm out}=2.25$ planet radii, similar to Saturn's ring 
\citep{planetary_science_Pater}. 
The ring optical thickness~$b$ is 1.0 and the ring particles have a single 
scattering albedo $\varpi$ of 0.8. 
Such a high albedo mimics the bright, icy particles in Saturn's ring. 
Considering the small probability that a ring around an observed 
exoplanet has either exactly 
$\gamma=0^\circ$ or $\gamma=90^\circ$ and $\lambda_{\rm r}=0^\circ$ 
or $\lambda_{\rm r}=90^\circ$, we use more arbitrary parameter values. 
For our standard system, $i=20^\circ$, $\lambda_{\rm r}=30^\circ$, 
and $\gamma=60^\circ$. The ring-plane crossings in this 
system happen at $\nu = 69.4^\circ$ and $\nu = 249.4^\circ$
(see Eq.~\ref{eq:ring_plane_crossing}). 
Between these values of $\nu$, the ring is seen in diffusely
transmitted starlight, while at smaller or 
larger values of $\nu$, the ring is seen in reflected starlight.

In Sect.~\ref{sec:ring_orientation}, we vary the ring orientation and 
orbital inclination. 
In Sects.~\ref{sec:results_optical_thickness} and
\ref{sec:results_albedo}, we show the influence of the 
ring optical thickness $b$ and the ring particle albedo $\varpi$.
And lastly, in Sect.~\ref{sec:results_ring_size}, 
we look at the effect of different values of the outer 
ring radius $r_{\rm out}$ on the light curves.

\subsection{Ring-less planet}
\label{sec:ring_less}
%%%%%%%%%%%%%%%%%%%%%%%%%%%%%%%%%%%%%%%%%%%%%%%%%%%%%%%%%%%%%%%%%
%%%%%%%%%%%%%%%%%%%%%%%%%%%%%%%%%%%%%%%%%%%%%%%%%%%%%%%%%%%%%%%%%
\begin{figure}
\includegraphics[width=9cm]{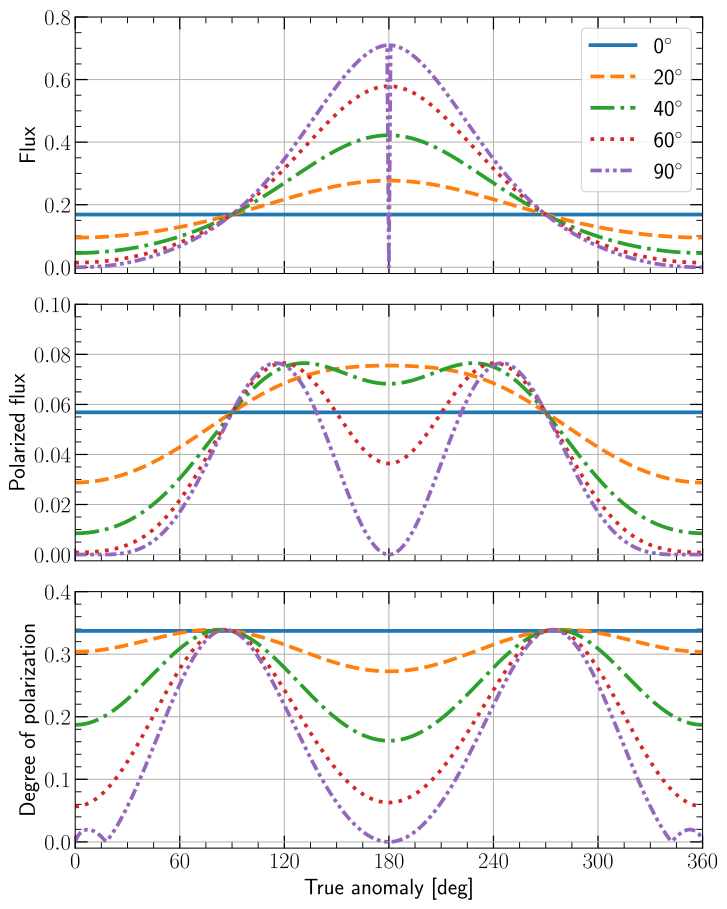}
\caption{The influence of the orbital inclination angle $i$ on the 
         reflected light of the ring-less planet as functions of the
         planet's true anomaly $\nu$.
         From top to bottom: 
         total flux $F$, polarized flux $F_{\rm pol}$, 
         and degree of polarization $P$. 
%         For $i=0^\circ$, phase angle $\alpha$ is 90$^\circ$ 
%         everywhere along the orbit. 
%         For $i > 0^\circ$, 
%         the planet is at its maximum phase angle $\alpha_{\rm max}$ 
%         at $\nu=0^\circ$ and 360$^\circ$, and at its minimum phase angle
%         $\alpha_{\rm min}$ at $\nu=180^\circ$ (for $i=90^\circ$, the 
%         planet is then precisely behind its star). 
         The different lines pertain to the following:
         $i=0^\circ$ (solid, blue): $\alpha= 90^\circ$;
         $i=20^\circ$ (dashed, orange): $\alpha_{\rm min}=70^\circ$,
         $\alpha_{\rm max}=110^\circ$;
         $i=40^\circ$ (dot-dashed, green): $\alpha_{\rm min}= 50^\circ$,
         $\alpha_{\rm max}=130^\circ$; 
         $i=60^\circ$ (dotted, pink): $\alpha_{\rm min}=30^\circ$,
         $\alpha_{\rm max}=150^\circ$;
         $i=90^\circ$ (dot-dashed, purple): $\alpha_{\rm min}=0^\circ$,
         $\alpha_{\rm max}=180^\circ$.
    }
\label{fig:Orbit_incl}
\end{figure}
%%%%%%%%%%%%%%%%%%%%%%%%%%%%%%%%%%%%%%%%%%%%%%%%%%%%%%%%%%%%%%%%%
%%%%%%%%%%%%%%%%%%%%%%%%%%%%%%%%%%%%%%%%%%%%%%%%%%%%%%%%%%%%%%%%%

\noindent Figure~\ref{fig:Orbit_incl} shows the reflected flux~$F$, 
the polarized flux~$F_{\rm pol}$, and the degree of 
polarization~$P$ of our model planet as functions of the  
true anomaly $\nu$ for orbital inclination angles $i$ ranging 
from 0$^\circ$ (a face-on orbit) to 90$^\circ$ (an edge-on orbit). 
The total and polarized fluxes are normalized and 
therefore unit-less (see Sect.~\ref{sec:integrated}). 
Similar curves have been 
presented by e.g.\ \citet{stam2004using,buenzli2009grid}. 

As can be seen in Fig~\ref{fig:Orbit_incl}, for $i=0^\circ$, 
phase angle $\alpha$ is always 90$^\circ$, and  
$F$, $F_{\rm pol}$, and $P$ are thus constant along the orbit. 
For $i > 0^\circ$, the planets attain their largest phase angles
at $\nu=0^\circ$ and 360$^\circ$, and their smallest phase
angle at $\nu=180^\circ$. For the edge-on orbit
($i=90^\circ$), the planet is precisely in front of its star
at $\nu=0^\circ$ and 360$^\circ$, and thus in transit, and 
at $\nu=180^\circ$, the planet is precisely behind its star.
While transit signals can be computed with {\Pryn}, they are
not included in our simulations. 

As expected, $P$ is largest around $\nu=90^\circ$ and $270^\circ$ 
when $\alpha \approx 90^\circ$ and the single scattering degree of 
polarization of the gaseous molecules is highest (see 
Fig.~\ref{fig:single_scattering}).
The peak polarized flux shifts towards $\nu=180^\circ$ 
with decreasing $i$ 
because it is modulated with the total amount of reflected 
light, which increases with decreasing $\alpha$. 
The small peaks in $P$ for $i=90^\circ$ and small and large values
of $\nu$ are caused by light that has been scattered twice 
\citep{stam2004using}. The direction of polarization of this 
twice scattered light is parallel to the detector plane.

%----------------------------------------------------------------------------
%----------------------------------------------------------------------------
\subsection{The influence of the ring orientation}
\label{sec:ring_orientation}

%%%%%%%%%%%%%%%%%%%%%%%%%%%%%%%%%%%%%%%%%%%%%%%%%%%%%%%%%%%%%%%%%
%%%%%%%%%%%%%%%%%%%%%%%%%%%%%%%%%%%%%%%%%%%%%%%%%%%%%%%%%%%%%%%%%
\begin{figure*}[ht]
\centering
\includegraphics[width=\textwidth]{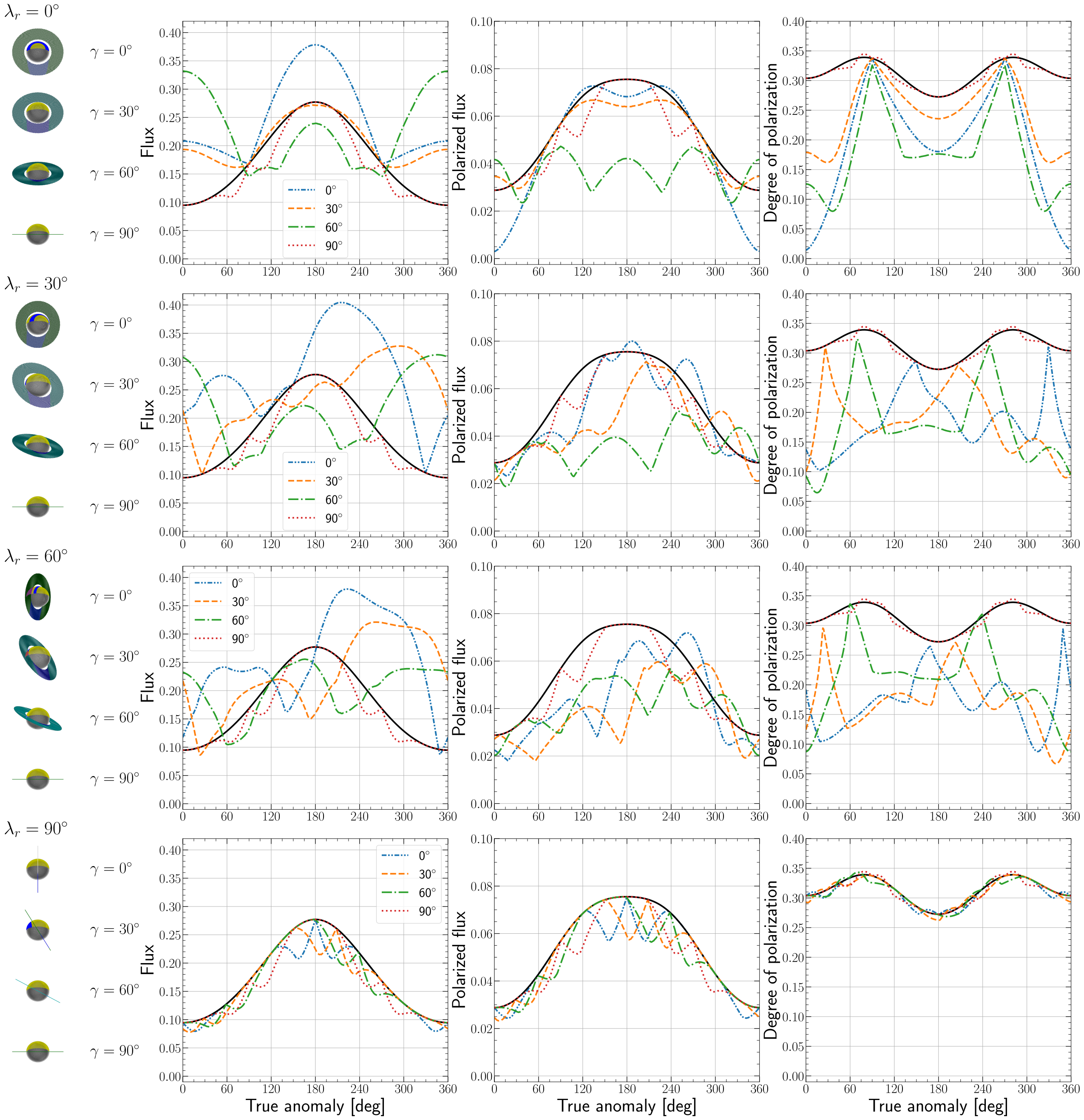}
\caption{$F$ (left), $F_{\rm pol}$ (middle), and $P$ (right) of the 
         light that is reflected by the model planet with a Saturn-like ring 
         with $r_{\rm in}=1.2$ and $r_{\rm out}=2.25$, $b=1.0$, and $i=20^\circ$
         ($\alpha_{\rm min}=70^\circ$ and $\alpha_{\rm max}=110^\circ$) as
         functions of the true anomaly $\nu$.
         The ring inclination longitude $\lambda_{\rm r}$ is 0$^\circ$ (first row),
         30$^\circ$ (second row), $60^\circ$ (third row), and 90$^\circ$ 
         (bottom row). For $\lambda_{\rm r}=90^\circ$ the ring is 'seen' edge-on. 
         The ring inclination angle $\gamma$ is 0$^\circ$
         (blue, dash-dot-dot), 30$^\circ$ (green, long-dashed),
         60$^\circ$ (red, long-dash-dot), or 90$^\circ$ (purple, dot-dot). 
         For $\gamma=90^\circ$, there is no dependence on $\lambda_{\rm r}$.
         The black lines are for the planet without a ring. 
         The images on the left illustrate the planet with its ring at 
         $\nu=0^\circ$, thus when it is closest to the observer.}
\label{fig:geom_orbit_i20}%
\end{figure*}
%%%%%%%%%%%%%%%%%%%%%%%%%%%%%%%%%%%%%%%%%%%%%%%%%%%%%%%%%%%%%%%%%
%%%%%%%%%%%%%%%%%%%%%%%%%%%%%%%%%%%%%%%%%%%%%%%%%%%%%%%%%%%%%%%%%

\noindent Here we show the influence of the orientation of the ring 
for a given planetary orbital inclination angle $i$. To limit the 
number of plots, 
we use only two values for $i$, namely $20^\circ$ and $90^\circ$, 
with the first value representing a planetary system 
that would have been discovered using direct imaging, and the 
latter a system typically discovered using the transit method. 

\begin{figure*}[ht]
\centering
\includegraphics[width=\textwidth]{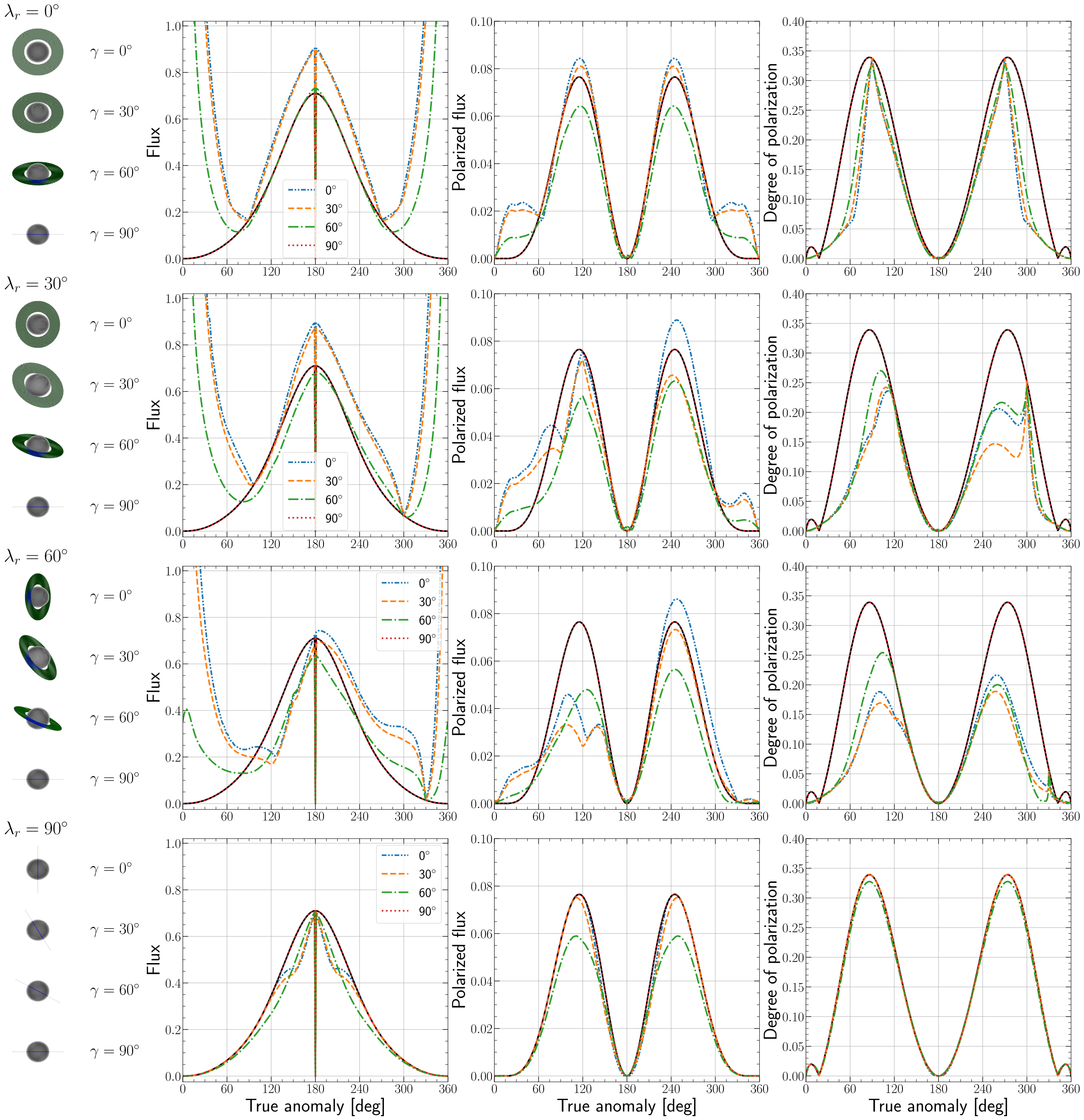}
\caption{Similar to Fig.~\ref{fig:geom_orbit_i20}, except for $i=90^\circ$. 
         The images on the left illustrate the system at $\nu=0^\circ$,
         which, for $i=90^\circ$, is precisely in front of the star,
         thus the nightside of the planet is turned towards the observer. 
         To keep the structures in the curves visible, we have limited the
         vertical axis in the graphs for $F$, thus in some cases cutting
         off the forward scattering peaks. The missing peak values are 
         the following: for $\lambda_{\rm r}=0^\circ$, the lines reach 
         6.13, 5.37, and 2.25 for $\gamma=0^\circ,\ 30^\circ$, 
         and $60^\circ$, respectively. For $\lambda_{\rm r}=30^\circ$,
         they reach 5.36, 4.61, and 1.55, and for $\lambda_{\rm r}=30^\circ$, 
         2.38 and 1.65 for $\gamma=0^\circ$ and $\gamma=30^\circ$, 
         respectively. }
\label{fig:geom_orbit_i90}
\end{figure*}

Figure~\ref{fig:geom_orbit_i20} shows for $i=20^\circ$, $F$, 
$F_{\rm pol}$, and $P$ for ring inclination angles $\gamma$ 
of 0$^\circ$ (a face-on ring), 30$^\circ$, 
60$^\circ$, and 90$^\circ$ (an edge-on ring), 
and for ring inclination longitudes $\lambda_{\rm r}$ of 
0$^\circ$, 30$^\circ$, 60$^\circ$, and 90$^\circ$. 
Figure~\ref{fig:geom_orbit_i90} is similar to 
Fig.~\ref{fig:geom_orbit_i20} except for $i=90^\circ$. 
For comparison, the figures also include lines 
representing the ring-less planet. 
Most of the curves in Figs.~\ref{fig:geom_orbit_i20} 
and~\ref{fig:geom_orbit_i90} show angular features that are due to 
changing shadows (cf.\ Fig.~\ref{fig:diff_positions}). 
We will not discuss all features in detail but rather point out a few
characteristic ones.

First, while the phase curves of the planet itself are symmetric 
around $\nu=0^\circ$, 
a non-zero ring inclination longitude $\lambda_{\rm r}$ makes them asymmetric. This has been noted before \citep{Arnold2004, dyudina2005phase}. 
% We first discuss the case of $i=20^\circ$ (Fig.~\ref{fig:geom_orbit_i20}). 
For a given $i$, all curves with $\gamma=90^\circ$ (an edge-on ring) are the same, because the view
on the ring is independent of $\lambda_{\rm r}$. In this edge-on orientation, the 
observer receives no light that has been reflected by or transmitted 
through the ring. However, the ring does leave traces in the light curves. 
For example, in Fig.~\ref{fig:geom_orbit_i20} ($i=20^\circ$, and $\gamma=90^\circ$) the shadow 
the ring casts on the planet reduces $F$ and $F_{\rm pol}$ 
between approximately $\nu=50^\circ$ and $130^\circ$, and $230^\circ$ and $310^\circ$. 
For $\lambda_{\rm r}=90^\circ$, the ring is seen edge-on for all $\gamma$ 
and any difference with the ring-less planet curves is thus due to ring shadows. 
It is interesting to note that very few geometries allow the 
ring to go completely undetected. One such geometry is shown in
Fig.~\ref{fig:geom_orbit_i90} ($i=90^\circ$. When $\gamma=90^\circ$ the thin ring 
does not cast a shadow nor does it reflect (or transmit) light towards 
the observer. In fact, any configuration where the orbital inclination 
equals the ring inclination angle leaves the ring undetectable\footnote{
If planet-shine were to be added to the simulations, the ring would theoretically only be
undetectable for $\gamma=90^\circ$ and $i=90^\circ$.}.

For all geometries, the curves of the ringed planets approach those of 
the ring-less planet at the ring-plane crossings, where the ring is 
illuminated on its edge and the shadow is infinitely narrow. 
Depending on its orientation, the ring can, however, still occult part of the
planetary disk at the ring-plane crossings, leaving small differences between
the curves, see for example in Fig.~\ref{fig:geom_orbit_i20}, 
the curve for $\lambda_{\rm r}=0^\circ$ and 
$\gamma=60^\circ$, where $F$ of the ringed planet is slightly lower than 
that of the ring-less planet at the ring-plane crossing. 
For $\lambda_{\rm r}=0^\circ$, the ring-plane crossings 
occur at $\nu=90^\circ$ and $270^\circ$ but for other values of $\lambda_{\rm r}$ 
and/or $\gamma$ (the latter only when $\lambda_{\rm r} \neq 0$), the locations of 
the ring-plane crossings are different. 

The ring-plane crossings usually manifest themselves in the 
light curves as sharp changes which can be described as 
a discontinuity in the derivative due to the light suddenly 
having to travel through the ring. A prime 
example of such a discontinuity in Fig.~\ref{fig:geom_orbit_i20} is the 
$\gamma=60^\circ$ curve at $\lambda_{\rm r}=0^\circ$. Between $\nu=0^\circ$ 
and $90^\circ$, the ring reflects light, strongly increasing $F$ 
of the planet-ring system compared to that of a ring-less planet. 
After the ring-plane crossing, the ring instead transmits light and casts 
a shadow on the planet, decreasing $F$ compared to the ring-less planet. 
The $\gamma=0^\circ$ curve in the same plot shows similar behavior but 
here the ring transmits and reflects along different parts of the orbit. 
The ring-plane crossings appear to be even more pronounced in $P$ than in
$F$ and $F_{\rm pol}$. 

Looking at the curves of $F_{\rm pol}$, it is clear that the light that 
is reflected or transmitted by the ring is usually less polarized than 
the light that is reflected by the planet. This suppresses $P$ of the 
system as a whole and is straightforward to understand from 
Fig.~\ref{fig:single_scattering}:
the light that is singly scattered by the ring particles not only has
a lower $P$, but at large scattering angles (small planetary phase angles 
$\alpha$) it also has an opposite direction of polarization compared 
to the light that is scattered in the planetary atmosphere, thus 
further decreasing $P$ of the whole.

The ring shadow on the planet can increase $P$ when it breaks the 
symmetry of the illuminated and visible planetary disk. 
Examples of this in Fig.~\ref{fig:geom_orbit_i20} are the curves with $\gamma=90^\circ$: 
the edge-on ring prevents it from reflecting light to the observer, 
but $P$ is slightly higher just before and after the ring-plane crossings. 
Another example in Fig.~\ref{fig:geom_orbit_i20} are the drops 
in the $P$-curves for $\gamma=60^\circ$ and 
$\lambda_{\rm r} = 0^\circ$ and $30^\circ$ before and after the ring-plane
crossings, where the ring casts its shadow on the planet.

The $F_{\rm pol}$-curves due to the ring are less 
straightforward to understand because they are not consistently 
lower than those of the ring-less planet. 
For example, in Fig.~\ref{fig:geom_orbit_i20} for $\lambda_{\rm r}=30^\circ$ 
and $\lambda_{\rm r}=60^\circ$, the ring adds $F_{\rm pol}$ in the second part of the orbit. 
This is due to the high flux of the ring there, which even though it is 
only weakly polarized, still adds polarized flux to the total $F_{\rm pol}$.
This effect is also visible in the $F_{\rm pol}$-curves in 
Fig.~\ref{fig:geom_orbit_i90} for $\lambda_{\rm r} < 90^\circ$
that show bumps at the beginning and end of the orbit.  

%%%%%%%%%%%%%%%%%%%%%%%%%%%%%%%%%%%%%%%%%%%%%%%%%%%%%%%%%%%%%%%%%
%%%%%%%%%%%%%%%%%%%%%%%%%%%%%%%%%%%%%%%%%%%%%%%%%%%%%%%%%%%%%%%%%
%%%%%%%%%%%%%%%%%%%%%%%%%%%%%%%%%%%%%%%%%%%%%%%%%%%%%%%%%%%%%%%%%
%%%%%%%%%%%%%%%%%%%%%%%%%%%%%%%%%%%%%%%%%%%%%%%%%%%%%%%%%%%%%%%%%

For $i=90^\circ$ in Fig.~\ref{fig:geom_orbit_i90}, the ring is much 
brighter in diffusely transmitted light than in reflected light
(note that the peak values of $F$ at small and large values of $\nu$
are mentioned in the figure caption). 
These peaks can be explained by the strong forward scattering 
peak in the single scattering flux of the ring particles  
(Fig.~\ref{fig:single_scattering}), especially in a part of the
orbit where the planet itself is dark.
It might seem that this forward scattering peak would easily be
detectable for a planet-ring system. However, in this part of the
orbit, the angular separation between the planet and its star is very 
small, which makes direct detections very hard. 
Such peaks could however be observable when monitoring the brightness
of the star and variations therein. This has already been demonstrated 
by \citep{placek2014exonest} in the infrared where hot giant planets 
are relatively bright but could possibly also be done at visible
wavelengths \citep{Sucerquia2020}. 

Changing the ring inclination longitude $\lambda_{\rm r}$ changes the 
locations of the ring-plane crossings but, unsurprisingly, not those 
of the forward scattering peak. 
The dramatic drops in $P$ that could be seen in 
Fig.~\ref{fig:geom_orbit_i20} are also present in 
Fig.~\ref{fig:geom_orbit_i90} but no longer significantly change the 
shape of the curves. Instead, the peak value of $P$ is often 
decreased, making it difficult to distinguish the curves from 
those of a ring-less planet. 
A ring-less planet with for example clouds could also significantly 
influence $P$ \citep[see][]{stam2004using,karalidi2012looking}. 
The asymmetry of the curves with $\lambda_{\rm r}=30^\circ$ could
possibly help to identify the rings, although it would have to be
ruled out that the asymmetrical behavior would be due to seasonal 
effects on a horizontally inhomogeneous ring-less planet as was 
remarked by \cite{dyudina2005phase}. 
Seasonal effects could, however, not explain the sharp changes due
to the ring-plane crossings, in $F$ and $P$ at, for example, 
$\nu=300^\circ$. Such sharp changes, which would require frequent observations of 
the system would be valuable characterizing features to search for.

%----------------------------------------------------------------------------
%----------------------------------------------------------------------------
\subsection{The influence of the ring optical thickness}
\label{sec:results_optical_thickness}

%%%%%%%%%%%%%%%%%%%%%%%%%%%%%%%%%%%%%%%%%%%%%%%%%%%%%%%%%%%%%%%%%
%%%%%%%%%%%%%%%%%%%%%%%%%%%%%%%%%%%%%%%%%%%%%%%%%%%%%%%%%%%%%%%%%
\begin{figure}
\includegraphics[width=9cm]{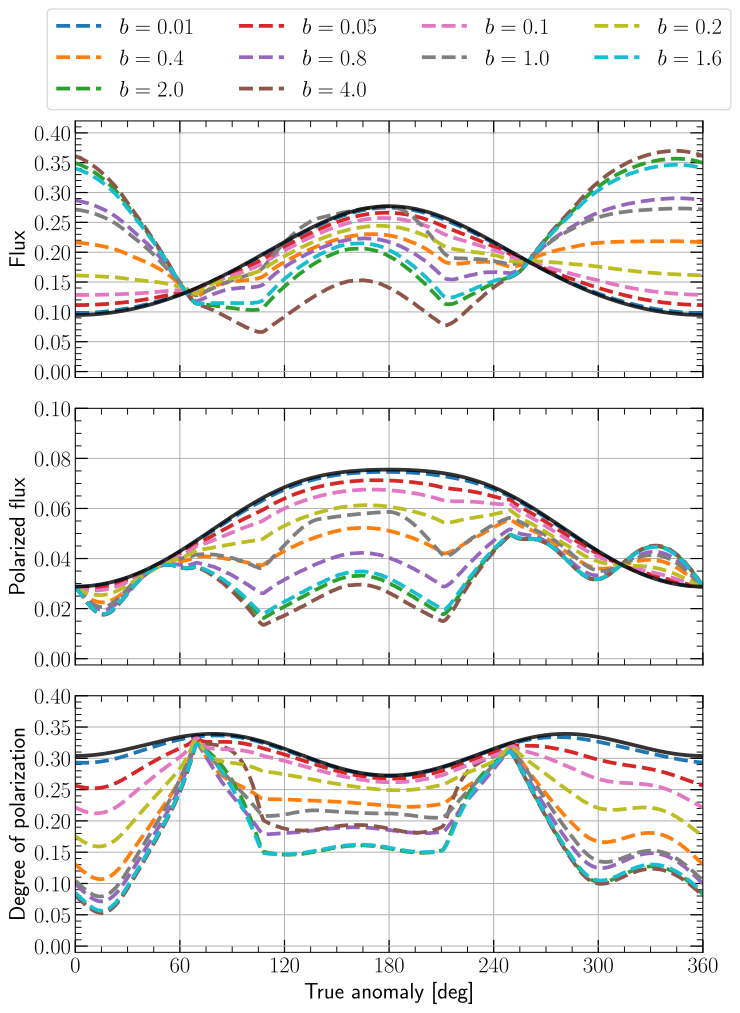}
\caption{$F$ (top), $F_{\rm pol}$ (middle), and 
         $P$ (bottom) as functions of $\nu$ for
         ring optical thicknesses $b$ ranging from 0.01 (dark blue)
         to 4.0 (brown).
         For this planet-ring system, $i=20^\circ$, 
         $\lambda_{\rm r}=30^\circ$, $\gamma=60^\circ$, $r_{\rm in}=1.20$, 
         $r_{\rm out}=2.25$, and $\varpi=0.8$. 
         The black line represents the ring-less planet.}
\label{fig:optical_sweep}
\end{figure}
%%%%%%%%%%%%%%%%%%%%%%%%%%%%%%%%%%%%%%%%%%%%%%%%%%%%%%%%%%%%%%%%%
%%%%%%%%%%%%%%%%%%%%%%%%%%%%%%%%%%%%%%%%%%%%%%%%%%%%%%%%%%%%%%%%%

Figure~\ref{fig:optical_sweep} shows the influence of the ring optical 
thickness~$b$ on the signals of the standard planet-ring system. 
First, we'll discuss the influence of $b$ along the
part of the orbit where the ring reflects incident light, 
and then the more complicated part where the ring diffusely 
transmits light.

In the figure, the ring reflects light for $\nu \leq 69.4^\circ$ 
or $\geq 249.4^\circ$ and with that generally increases $F$ of the system 
as a whole. The increase in $F$ increases with $b$, although not linearly. 
With increasing $b$, the increase in $F$ vanishes as the reflection 
by the ring reaches an asymptotic value when increasing $b$ does not
further increase $F$.
Note that along this part of the orbit, the planet also casts
its shadow on the ring, thus decreasing the ring-contribution to $F$,
while the ring hardly casts a shadow on the planet.

In reflected light, polarized flux $F_{\rm pol}$ of the system is 
lower than that of the ring-less planet at small
values of $\nu$, while it is larger at large values of $\nu$,
with the difference increasing with $b$ and converging for 
the larger values of $b$.
The reason for the lower and higher values of $F_{\rm pol}$ 
when the ring is added
are due to the polarized flux of the light that is singly scattered by the 
ring particles,
which has an opposite direction of that of the gas molecules in the 
planet's atmosphere at the small
values of $\nu$, where the single scattering angle $\Theta$ is larger than 
150$^\circ$, and the same direction at the large values of $\nu$, 
where $\Theta$ is smaller than 150$^\circ$
(see Fig.~\ref{fig:single_scattering}).

\begin{figure}
\includegraphics[width=9cm]{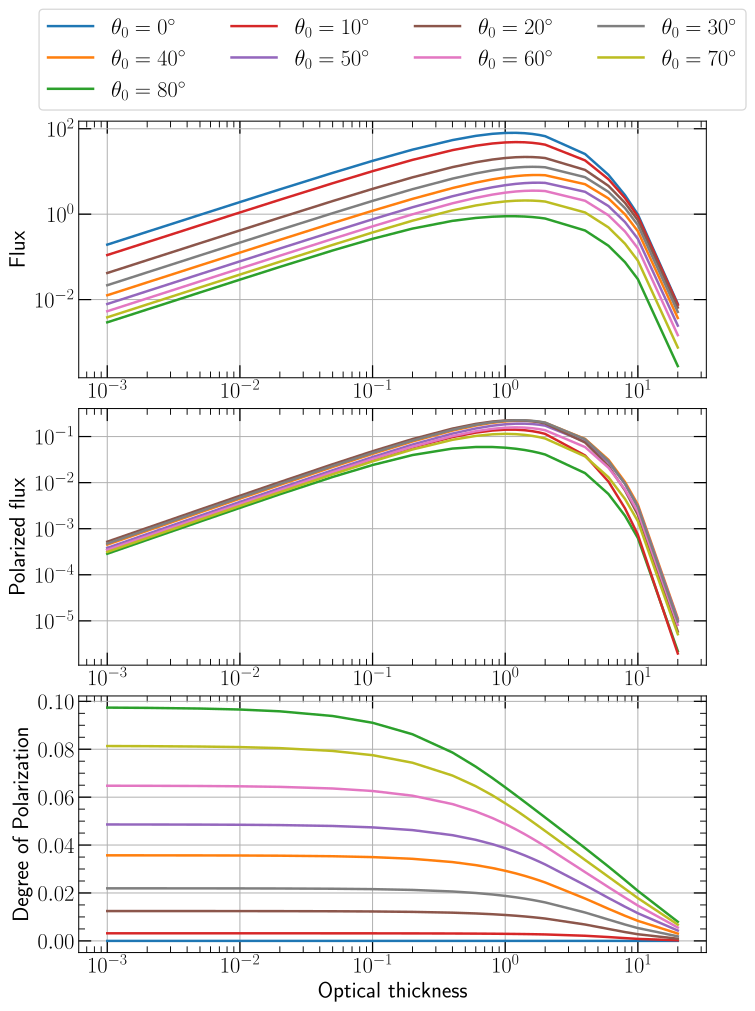}
\caption{The diffusely transmitted $F$ (top), 
         $F_{\rm pol}$ (middle), and $P$ (bottom) as functions of 
         the optical thickness $b$ of a slab of ring-material 
         for illumination angles $\theta_0$ 
         ranging from 0$^\circ$ (blue) to 80$^\circ$ (green). 
         The viewing angle $\theta$ is $0^\circ$.}
\label{fig:optical_vs_flux}
\end{figure}

Between $69.4^\circ < \nu < 249.4^\circ$, the ring diffusely 
transmits light, adding to the total flux of the system. At the same
time, the ring casts shadows on the planetary disk en occults part 
of the disk, both effects suppressing the planetary flux
(see Fig.~\ref{fig:diff_positions}). 
To understand the relation between the transmitted light and $b$,
Fig.~\ref{fig:optical_vs_flux} shows $F$, $F_{\rm pol}$, and $P$ of 
diffusely transmitted light of just a slab of ring-material 
for a viewing angle of $0^\circ$, 
different angles of incidence, as functions of $b$. 
As can be seen, the curves for $F$ increase with $b$ due to 
increased scattering of light by the ring particles (the flux of the 
directly, thus non-scattered beam of light decreases with increasing $b$), 
until about $b= 1$ to 3 (depending on $\theta_0$ and $\theta$ as those angles
affect the effective ring optical thickness). Further increases
of $b$ decrease the diffusely transmitted $F$ as fewer and 
fewer photons get through the ring.

The relative large values of $F$ between $\nu \approx 100^\circ$ 
and~$210^\circ$ in Fig.~\ref{fig:optical_sweep},
are due to the changing size of the planet shadow on the ring.
Similar behavior can be seen for $F_{\rm pol}$. 

The degree of polarization $P$ generally decreases with increasing 
$b$ as the amount of multiple scattered, generally low polarized, 
light, increases with $b$. 
Between $\nu=69.4^\circ$ and $249.4^\circ$ where the flux 
of the reflected ring-light is small,
this depolarizing effect is not very prominent, especially 
not for $b=4.0$: the very small ring-flux hardly contributes 
to $F$ and $F_{\rm pol}$ while at the same time, the ring
strongly shadows and occults the planet and thus also decreases
the total flux of the system.

%%%%%%%%%%%%%%%%%%%%%%%%%%%%%%%%%%%%%%%%%%%%%%%%%%%%%%%%%%%%%%%%%
%%%%%%%%%%%%%%%%%%%%%%%%%%%%%%%%%%%%%%%%%%%%%%%%%%%%%%%%%%%%%%%%%
%%%%%%%%%%%%%%%%%%%%%%%%%%%%%%%%%%%%%%%%%%%%%%%%%%%%%%%%%%%%%%%%%
%%%%%%%%%%%%%%%%%%%%%%%%%%%%%%%%%%%%%%%%%%%%%%%%%%%%%%%%%%%%%%%%%

%----------------------------------------------------------------------------
%----------------------------------------------------------------------------
\subsection{The influence of the single scattering albedo}
\label{sec:results_albedo}

Next, we vary the single scattering albedo $\varpi$ of the ring particles 
to mimic different particle compositions. The icy particles 
surrounding Saturn would, 
for example, not survive at the distance between the Earth and the Sun. 
Considering the recent discoveries of puffed-up planets, of which the 
transit depth combined with their mass would indicate very small densities, 
that could possibly be explained by the presence of a ring \citep{piro2020exploring}
as that could increase the transit depth, it is important to also look 
at refractory materials (which have a lower albedo in the visible) 
\citep{piironen1998albedo,ostrowski2019physical}. 

%%%%%%%%%%%%%%%%%%%%%%%%%%%%%%%%%%%%%%%%%%%%%%%%%%%%%%%%%%%%%%%%%
%%%%%%%%%%%%%%%%%%%%%%%%%%%%%%%%%%%%%%%%%%%%%%%%%%%%%%%%%%%%%%%%%
\begin{figure}
\centering
\includegraphics[width=9cm]{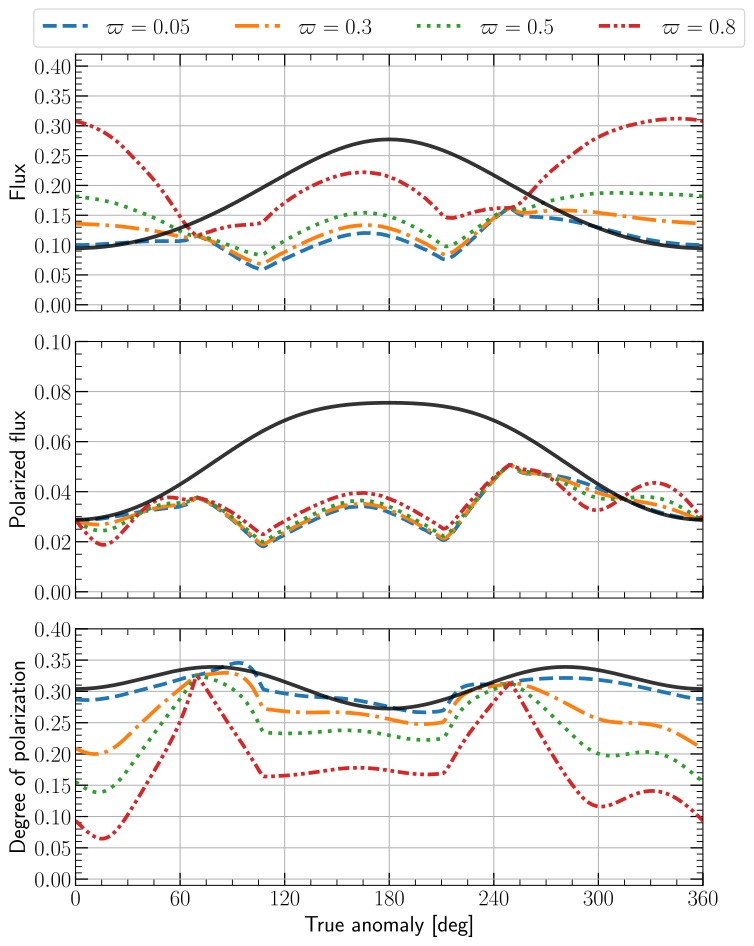}
\caption{Similar to Fig.~\ref{fig:optical_sweep}, except for 
         different single scattering albedos $\varpi$ 
         of the ring particles: 0.05 (blue, dashed),
         0.3 (orange, dashed-dot-dot), 0.5 (green, dotted), and
         0.8 (red, short-dashed-dotted).
         The system-parameters are $i=20^\circ$, $\lambda_{\rm r}=30^\circ$, 
         $\gamma=60^\circ$, $r_{\rm in}=1.20$, $r_{\rm out}=2.25$, and 
         $b=1.0$. }
\label{fig:albedo_sweep}
\end{figure}
%%%%%%%%%%%%%%%%%%%%%%%%%%%%%%%%%%%%%%%%%%%%%%%%%%%%%%%%%%%%%%%%%
%%%%%%%%%%%%%%%%%%%%%%%%%%%%%%%%%%%%%%%%%%%%%%%%%%%%%%%%%%%%%%%%%

Figure~\ref{fig:albedo_sweep} shows $F$, $F_{\rm pol}$, and $P$ for the
standard planet-ring system and $\varpi$ ranging from 0.05 to 0.8.
Not surprisingly, increasing $\varpi$ increases $F$, regardless of whether 
the ring is seen in reflected or transmitted light. Because for a given
$b$, $\varpi$ has no effect on the shadowing or occultation by the ring, 
and because the single scattering polarization of the particles is 
independent of $\varpi$, $F_{\rm pol}$ shows little dependence on 
$\varpi$, indirectly showing how small the contribution of the ring 
is to $F_{\rm pol}$ of the system. 
The variation in $P$ is thus mostly due to the variation in $F$.  

From comparing Figs.~\ref{fig:optical_sweep} and~\ref{fig:albedo_sweep},
we conclude that for this planet-ring system, it is
difficult to determine whether a curve would be due to a larger 
value of $b$ or a higher value of $\varpi$. 
Increases in either parameter lead to higher fluxes when the ring is 
reflecting light and a larger $b$ can also increase the transmitted flux.
For a different orientation of the ring and/or planetary orbit, this might 
be different, however, and that is something that should be studied 
with a full retrieval algorithm applied to simulated curves.

%----------------------------------------------------------------------------
%----------------------------------------------------------------------------
\subsection{The influence of the ring radius}
\label{sec:results_ring_size}

%%%%%%%%%%%%%%%%%%%%%%%%%%%%%%%%%%%%%%%%%%%%%%%%%%%%%%%%%%%%%%%%%
%%%%%%%%%%%%%%%%%%%%%%%%%%%%%%%%%%%%%%%%%%%%%%%%%%%%%%%%%%%%%%%%%
\begin{figure}
\includegraphics[width=9cm]{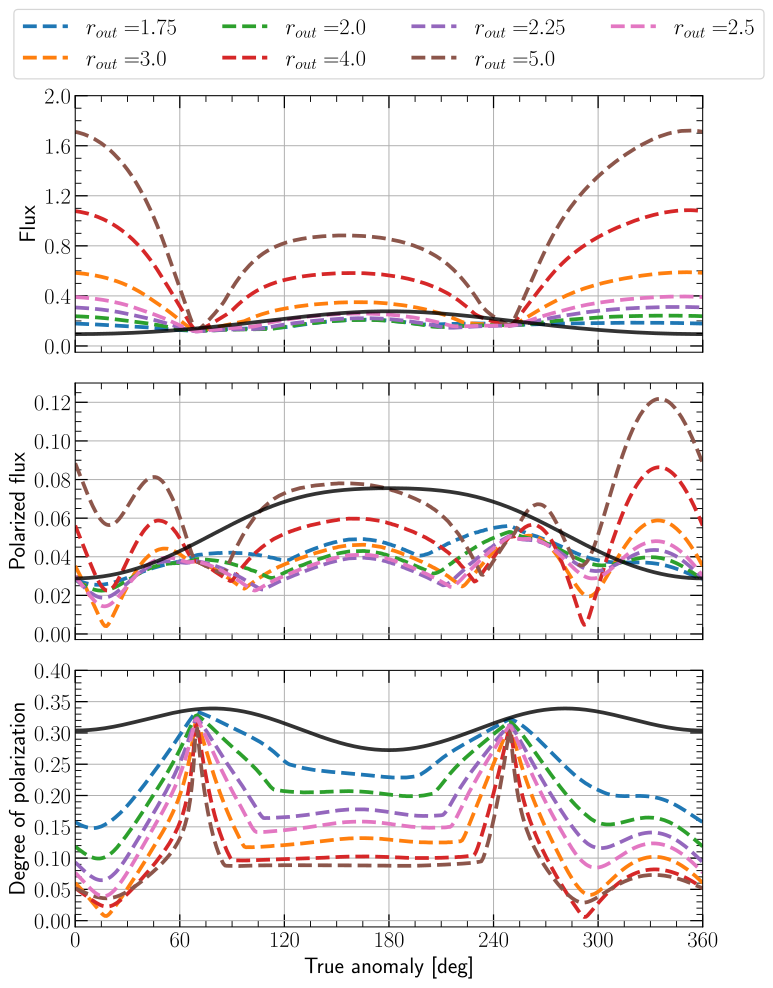}
\caption{Similar to Fig.~\ref{fig:optical_sweep}, except for 
         different outer ring radii $r_{\rm out}$ (expressed in 
         planet-radius $r$):
         1.75 (blue),
         2.0 (green), 2.25 (purple), 2.5 (pink),
         3.0 (orange), 4.0 (red), and 5.0 (brown).
         The system-parameters are $i=20^\circ$, $\lambda_{\rm r}=30^\circ$, 
         $\gamma=60^\circ$, $r_{\rm in}=1.2$, $\varpi=0.8$, and 
         $b=1.0$.}
   \label{fig:ring_size_sweep}
\end{figure}
%%%%%%%%%%%%%%%%%%%%%%%%%%%%%%%%%%%%%%%%%%%%%%%%%%%%%%%%%%%%%%%%%
%%%%%%%%%%%%%%%%%%%%%%%%%%%%%%%%%%%%%%%%%%%%%%%%%%%%%%%%%%%%%%%%%

The influence of the outer ring radius~$r_{\rm out}$ is shown in 
Fig.~\ref{fig:ring_size_sweep}. The curves are very similar to those 
for various ring optical thicknesses $b$ 
(Fig.~\ref{fig:optical_sweep}) since increasing $r_{\rm out}$ increases the 
reflected fluxes and, depending on the geometry, also the shadows in a 
similar way as increasing $b$ does. 
There are some key differences, however. For example, along parts of the 
orbit where the ring is diffusely transmitting light, the ring flux 
increases with 
increasing $r_{\rm out}$ while it would decrease with increasing $b$ 
beyond a value of about~1.0 (see Fig.~\ref{fig:optical_vs_flux}).

Another difference compared to changing $b$ is found in the 
curves for $F_{\rm pol}$: while increasing $b$ decreases $F_{\rm pol}$ 
of the system as a whole,
because light that has been scattered multiple times is less polarized, 
there is no such relation with the ring size:
the larger $r_{\rm out}$, the larger the $F_{\rm pol}$ that is added 
to the signal of the planet-ring system.
With increasing $r_{\rm out}$, the curves for $P$ converge as the polarization
signal of the rings starts to dominate that of the planet (except near the 
ring-plane crossings). 
This trend is helped by the fact that eventually, increasing $r_{\rm out}$ 
no longer increases the extent of the shadows on the planet.

The dips around $\nu=15^\circ$ and $290^\circ$ become more pronounced 
when the ring brightens as is evident from 
Figs.~\ref{fig:optical_sweep},~\ref{fig:albedo_sweep},~and~\ref{fig:ring_size_sweep}. 
At these locations in the orbit, the angle of polarization of the ring-light 
is opposite to that of the planet-light, thus decreasing $P$; an effect that
becomes more prominent with increasing $r_{\rm out}$. 
For $r_{\rm out}=5.0$, the flux that is reflected by the ring dominates the 
planetary flux and $P$ no longer approaches zero.

%----------------------------------------------------------------------------
%----------------------------------------------------------------------------
\section{The case of HIP~41378~f}
\label{sec:case_study}

As a case study, we simulate the flux and polarization 
signals of ``puffed-up'' planet HIP~41378~f 
\citep{akinsanmi2020planetary,alam2022first}, assuming that it is not a
single planet, but a planet with a ring.
HIP~41378~f is suspected to have a ring because of its extremely 
small average density of $0.09 \pm 0.02$~g~cm$^{-3}$ that follows from 
fitting the transit with a ring-less planet. A follow-up study by 
\cite{alam2022first} showed that the hypothesis of a ring holds up when 
the observations are done over a range of wavelengths but so does the hypothesis of 
an extended atmosphere. Assuming that the transit depth is partly caused by a ring, 
\cite{akinsanmi2020planetary} find an average density $\rho_{\rm p}$ of
$1.2 \pm 0.4$~g~cm$^{-3}$ instead. 

HIP~41378~f has a relatively large semi-major orbital axis of about
1.4~AU that, at a distance of 103~pc, translates to a sky-projected 
angular separation with its F-type parent star of $\sim$13~mas
\citep{santerne2019extremely}.
This would be large enough to resolve the planet by direct imaging 
telescopes like the proposed Large UV/Optical/Infrared Surveyor (LUVOIR) 
\citep{LUVOIR2019LUVOIR}. 
The planet has an equilibrium temperature $T_{\rm eq}$ of $\sim$294~K 
(assuming a bond albedo of zero, \citealt{santerne2019extremely}),
so there will be no significant thermal emission that would mix in
with the reflected starlight and decrease the planet's degree of 
polarization $P$. 
The transit of HIP~41378~f has been observed as part of the K2-mission 
\citep{howell2014k2mission} which observed at visible wavelengths 
that are comparable to the 633~nm that we used for our previous
computations.

%%%%%%%%%%%%%%%%%%%%%%%%%%%%%%%%%%%%%%%%%%%%%%%%%%%%%%%%%%%%%%%%%
%%%%%%%%%%%%%%%%%%%%%%%%%%%%%%%%%%%%%%%%%%%%%%%%%%%%%%%%%%%%%%%%%
\begin{table}
\begin{center}
\caption{HIP 41378~f parameters (without uncertainties)
         as derived by \cite{akinsanmi2020planetary}. 
         The orientation of the ring is defined differently by
         \cite{akinsanmi2020planetary} so our values for $\gamma$ and
         $\lambda_{\rm r}$ are different than theirs for the 
         same ring orientation. $r_{\rm p}$ is the planet radius
         (expressed in Earth radius $r_{\oplus}$), $a$ is the
         orbital semi-major axis, $r_*$ is the stellar radius.
         % {\color{red}{or of the Sun?}}
         }
\begin{tabular}{ll}
\hline
Parameter \hspace*{1cm} & Value \\ \hline
$r_{\rm p}$ [$r_{\oplus}$] & 3.7 \\
$a/r_*$          & 231.0 \\
$i$ [$^\circ$]  & 89.97 \\
$r_{\rm in}$ [$r_{\rm p}$] & 1.05 \\
$r_{\rm out}$ [$r_{\rm p}$] & 2.6 \\
$\gamma$ [$^\circ$] & -2.11 \\
$\lambda_{\rm r}$ [$^\circ$]  & -24.92 \\
\hline
\end{tabular}
\label{tab:HIPparameters}
\end{center}
\end{table}
%%%%%%%%%%%%%%%%%%%%%%%%%%%%%%%%%%%%%%%%%%%%%%%%%%%%%%%%%%%%%%%%%
%%%%%%%%%%%%%%%%%%%%%%%%%%%%%%%%%%%%%%%%%%%%%%%%%%%%%%%%%%%%%%%%%

A detailed fit of the transit using {\tt Pryngles} is outside 
the scope of this study. Instead, the results presented in this
section demonstrate the capabilities of {\tt Pryngles} and provide
an estimate of the detectability of the reflected light signals 
of this system. 
The parameters of a ringed planet that were determined by 
\citep{akinsanmi2020planetary} are shown in Table~\ref{tab:HIPparameters}. 
In order to show the detectability of the reflected light, 
we will use extremes of the ring optical thickness $b$ 
and the single scattering albedo $\varpi$ of the ring particles 
that are possible within the constraints of the values in 
Table~\ref{tab:HIPparameters}. 
Especially $\varpi$ is unconstrained by the transit but 
can have significant influence on the reflected flux
(see Fig.~\ref{fig:albedo_sweep}).
We will use values of $\varpi$ equal to 0.05 and 0.8 (see 
Fig.~\ref{fig:albedo_sweep}) where we note that a value of 0.8 
is high for the ring particles as they presumably consist of 
rocky material considering the proximity to the star. 
Both \citet{akinsanmi2020planetary,alam2022first}, however,
derive an average density $\rho$ of the ring particles of $1.08 \pm 0.3 
$~g~cm$^{-3}$, which is low compared to most rocky materials. 
This can be explained by porous materials \citep{carry2012density} 
or ice that is continuously being deposited by an unknown source.

Furthermore, both \citet{akinsanmi2020planetary} and \citep{alam2022first}
assume an opaque ring. While this assumption is not realistic, 
model simulations show that for $b \geq 4$, the transit depth would
be fairly close to the observations. Therefore, we vary $b$ between
4 and 20, large enough values to be considered an opaque ring. 
The orientation and ring radius are a much more direct result of 
the transit fit, and we'll thus adopt the values derived from the
fit (Table~\ref{tab:HIPparameters}). 
We again use the scattering properties of the olivine particles
that are comparable in size to the particles that were used in 
the follow-up study by \citep{alam2022first}. 
For simplicity, we use the same gaseous planetary atmosphere as 
before. 

%%%%%%%%%%%%%%%%%%%%%%%%%%%%%%%%%%%%%%%%%%%%%%%%%%%%%%%%%%%%%%%%%
%%%%%%%%%%%%%%%%%%%%%%%%%%%%%%%%%%%%%%%%%%%%%%%%%%%%%%%%%%%%%%%%%
\begin{figure}
   \includegraphics[width=9cm]{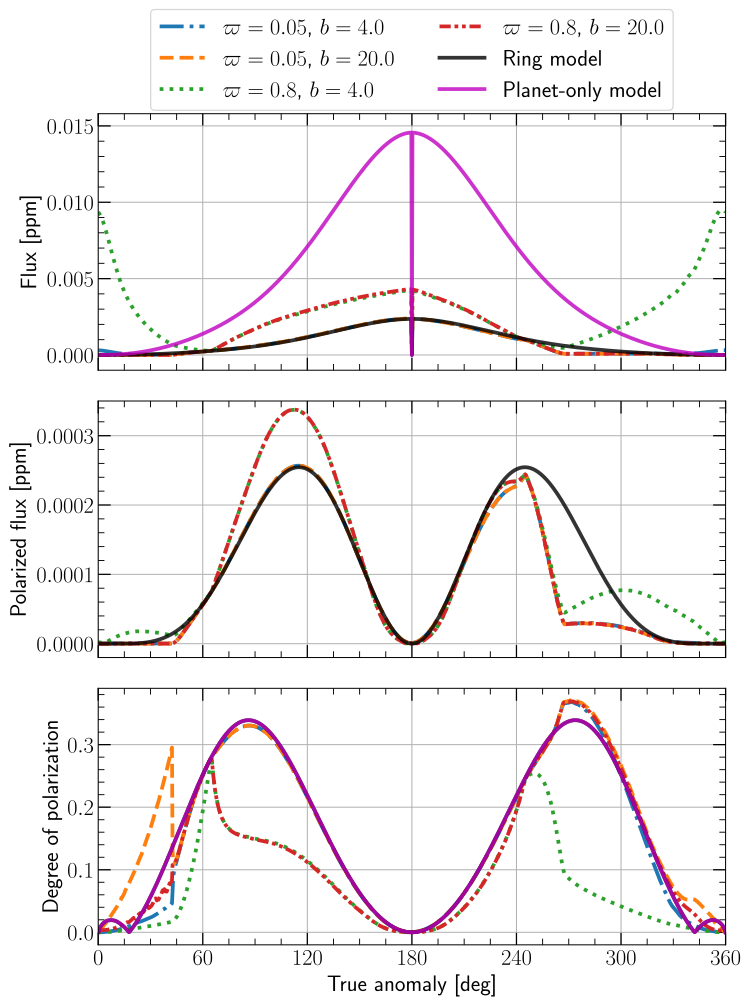}
   \caption{$F$, $F_{\rm pol}$, and $P$ of HIP 41378f for different values 
            of $b$ and $\varpi$ as functions of $\nu$. 
            The magenta curves pertain to the ring-less, "puffed-up" 
            planet and the black curves to the ringed-planet model but without a ring,
            using the parameter values of \cite{akinsanmi2020planetary}. 
            All curves use the same orbital inclination and stellar properties but 
            the magenta curve has a different semi-major axis ($a/r_* = 231.6$).
            In the $F_{\rm pol}$-plot, the magenta curve has been 
            omitted as it has the same shape as the black curve except 
            with maximum values up to $~$0.0015~ppm. 
            In the $P$-plot, the magenta and black curves overlap.
            % {\color{red}{what is the geometry for the black curve?}}
            }
\label{fig:HIP_tau_sweep}
\end{figure}
%%%%%%%%%%%%%%%%%%%%%%%%%%%%%%%%%%%%%%%%%%%%%%%%%%%%%%%%%%%%%%%%%
%%%%%%%%%%%%%%%%%%%%%%%%%%%%%%%%%%%%%%%%%%%%%%%%%%%%%%%%%%%%%%%%%

Figure~\ref{fig:HIP_tau_sweep} shows $F$, $F_{\rm pol}$, and $P$
for the various model planets and rings. It can be seen that 
even in the most favorable case, the contrast required to 
detect the planet with rings is on the order of $10^{-9}$. 
This is not attainable with SPHERE/ZIMPOL \citep{thalmann2008spherezimpol} 
or even with JWST \citep{carter2022JWST}, especially at the small 
angular separation for the maximum flux of 13~mas. 
If this contrast and resolution can be achieved in the 
near-future, we would be able to distinguish between a ring-less
or a ringed planet. 
Note that the large spike in $P$ before $\nu=60^\circ$ would not be 
observable due to the low $F$ at that location in the orbit. 

If we take the curve for the planet-only model as unknown in exact brightness but having the presented shape, then it becomes clear that at least two measurements are necessary to reliably differentiate between the two models. Considering the angular separation is largest around $\nu=90^\circ$ and $270^\circ$ the focus should be on differences in that region. Luckily, this is also where the biggest differences are in both $F$ and $P$. 
If there is a ring, a flux difference would be found when measuring at $\nu=90^\circ$ and $270^\circ$ which would not be the case for the ring-less planet. This flux difference is there regardless of the optical thickness or single scattering albedo of the ring but could be absent or changed if the ring has a different orientation. If the polarization of the light is also measured, something could be said about the composition of the ring. For example, assuming that the ring is truly opaque, so $b=20$, $P$ at $\nu=90^\circ$ would be a function of $\varpi$ and so could potentially be extracted. A thorough analysis of observations could extract both the optical thickness and albedo based on the difference in $P$ at $\nu=90^\circ$ and $270^\circ$. 

Based on these results it should be possible to distinguish between a planet with a simple atmosphere and a planet with a ring if they could be directly observed. If $P$ is also measured it should even be possible to extract some of the properties of the ring. As was mentioned in Sect.~\ref{sec:ring_orientation} and by \cite{dyudina2005phase}, a planet that shows seasonal activities might also produce asymmetrical light curves. A follow-up study could investigate this possibility based on the proposed atmospheres by \cite{alam2022first}. A third possible explanation comes from a recent study that shows that the observed transit can be explained by the presence of an exomoon \citep{harada2023stability}. The reflected light of a planet with an exomoon could in theory also lead to a measured flux difference because the moon may move in front of the planet, or vice versa. The chance of this happening during an observation depends on the size and period of the moon as well as the observation frequency but should be small, as was shown by \cite{molina2018traces}. They also showed that such transits (of the moon in front of the planet or vice versa) only result in shallow, brief dips in the light curve. The dips were especially small in the curves for $P$ (around $2\%$) and so should be distinguishable from the effects of a ring.

%--------------------------------------------------------------------------------
%--------------------------------------------------------------------------------
%--------------------------------------------------------------------------------
\section{Discussion}
\label{sec:discussion}
The work presented here provides significant enhancements to the current state of the art in light scattering and polarization modeling compared to \citealt{Lietzow2023}. Our model uses a scattering matrix based on measurements of real particles and, perhaps more importantly, is a publicly available package {\tt Pryngles}. Furthermore, our study explores different orbit orientations and considers the case of an actual planet, HIP 41378 f, which has an extremely low density, making it a prime candidate for reanalysis under the ringed planet model \citep{Zuluaga2015, akinsanmi2020planetary}. The particles that we used have more realistic scattering behavior than similarly sized spherical particles but also have properties that are not found in ring particles, such as the material that they are made of. For example, the rings of Saturn are mainly composed of icy materials \citep{cuzzi1984saturns} and have a different refractive index compared to the olivine particles that were used. In future updates, different ring particles will be added to the package based on data in the Amsterdam-Granada database \citep{munoz2012amsterdam-granada}. Another potential problem is that the used particles have a log-normal distribution \citep{munoz2000experimental,moreno2006scattering} while ring particles have a power-law distribution \citep{dohnanyi1969collisional,cuzzi1978saturn}. The difference in scattering behavior of the two different size distributions should be investigated in a future study. The effect of non-homogeneous rings should also be explored in a future study as our model ring is horizontally homogeneous, while real planetary rings, such as those in the solar system, are not. The rings in our solar system exhibit azimuthal variations, but especially radial variations in their optical thickness and particle properties. Such inhomogeneities will influence the ring shadows and occultations, and their traces in the signals of planet-ring systems would be interesting to explore.

In general, the developments presented here can contribute to three fronts in the search for exoplanets with rings: rings around exoplanets in close-in orbits (i.e. "warm exorings"), circumplanetary disks (CPDs), and rings around distant giant planets similar to Saturn's rings. With respect to the first case, populations of warm rings, proposed originally by \citealt{Schlichting2011}, can exist in proximity to their host stars for millions of years and will be composed of refractory particles rather than icy particles as in Saturn's rings. The thermal emission of warm rings cannot be modeled using our model but when computing flux or the degree of polarization, it could be added as a constant flux because the thermal emission is predicted to have a relatively low degree of polarization \citep{Stolker2017}. Because our scattering model as presented here is based on irregular olivine particles which, while they may not necessarily be found in rings, can represent refractory particles, our model can be used to model the reflected light.

In the second case, massive planets with extensive orbits have a higher likelihood of developing substantial circumplanetary disks, which may ultimately transform into planetary rings upon completion of the satellite formation process. The detection of such planet-ring systems could be achieved using high-contrast imaging techniques, which are the sole method of planet detection that offers spatial resolution of photons emanating from planets and their surroundings \citep[See, e.g.][]{Ruane2018}. This approach utilizes coronagraphs or other nulling methods to suppress light from the central star, and in conjunction with observing techniques such as Angular Differential Imaging (ADI), can attain contrasts of 10$^{-6}$ at separations of tenths of arcseconds \citep{Marois2010}. Consequently, this method enables the detection of self-luminous gas giants (i.e. planets that are still forming) on wide orbits and the effects of their presumed rings. To date, planets identified using direct imaging are generally a few million years old and situated tens of astronomical units from their host star favoring the survival of icy rings. High-contrast imaging has yielded several significant discoveries, including the detection of four planets orbiting HR 8799 \citep{Marois2010}, the planet $\beta$ Pic b \citep{Lagrange2010}, and others. As a result, the advancements presented in this paper demonstrate a potential for modeling the circumplanetary ring populations discovered through the aforementioned techniques.

%%PARAGRAPH BROUGHT FROM SECTION 2
Although {\Pryn} can handle eccentric orbits, we assume circular orbits to reduce the number of free parameters. Also, our results can be scaled for eccentric orbits by multiplying the total and polarized fluxes by a factor to represent the actual incident fluxes on the planet and its ring (the degree of polarization of the reflected signal would be unaffected by the orbital eccentricity). A potential complexity with simulating a ringed planet in an orbit with a large eccentricity is that depending on the size of the rings and the semi-major axis of the orbit, the illumination angle might no longer be uniform across the ring, while we assume a single illumination angle across the ring.\footnote{{\Pryn} itself can handle different illumination angles for different ring spangles.}

%%PARAGRAPH BROUGHT FROM THE INTRODUCTION
Our numerical predictions could not be compared against real observations of an actual ringed planet such as Saturn, because ground-based telescopes (or space telescopes in the vicinity of Earth), can only observe these planets at small phase angles, where the degree of polarization of the reflected light is very small because of a symmetric geometry. Observing these planets as if they were exoplanets is only possible with an orbiter, such as the Cassini spacecraft that orbited Saturn, or a fly-by mission. However, although Cassini's Imaging Science Subsystem (ISS) instrument had polarimetric capabilities \citep{porco2004Cassini}, no polarimetric observations of Saturn and its ring system have been published as of yet \citep{west2022personal}.

%%PARAGRAPH BROUGHT FROM THE INTRODUCTION
Polarimetry could also be a valuable method for detecting and characterizing the magnetic fields of ringed exoplanets. As in the case of the giant planets in the Solar System (and in the debris disks around young stars), a planet's magnetic field can modify the optical properties of the rings by influencing the particle orientations \citep{Dollfus1984, Lazarian2007}. Finding ring particle orientations could thus potentially provide valuable information about the structure and dynamics of the planet's interior, which in turn could provide clues about the habitability of these worlds (in case of rocky planets) and their satellites (see, e.g. \citealt{Heller2013}, and references therein).

Lastly, the study of light scattered from planetary rings around exoplanets is highly relevant to the current search for Saturn-like planets, and their characterization can reveal important information about their evolution and their environments. The modeling of the scattering and polarization of dust grains in planetary rings can provide valuable insights into the composition and evolution of the rings themselves, as well as the dynamics of the planet-ring system as a whole. Our study contributes to this field by providing a more accurate and realistic scattering model, which can be used in conjunction with high-contrast imaging techniques to search for and characterize exoplanet ring systems. By improving our understanding of the optical properties of planetary rings, we can gain further insights into the formation and evolution of exoplanet systems, and potentially identify new targets for future and ongoing missions such as the LUVOIR and the James Webb Space Telescope.

%--------------------------------------------------------------------------------
%--------------------------------------------------------------------------------
%--------------------------------------------------------------------------------
\section{Summary and Conclusions}
\label{sec:summary_conclusion}

We have computed the effects of a ring around an extrasolar planet 
on the total and polarized fluxes of starlight that is reflected by 
the planet-ring system as a whole. 
For these computations, we extended the python package {\Pryn} 
with an adding-doubling radiative transfer algorithm
that includes all orders of scattering and polarization 
\citep{haan1987adding}. 
We assumed dusty rings that are comprised of irregularly shaped 
particles with an effective radius of 1~$\mu$m. By varying 
system parameters, we investigated their role on the system's flux and polarization light curves. 
The system parameters that we varied are the ring orientation 
(Sect.~\ref{sec:ring_orientation}), the ring optical thickness 
(Sect.~\ref{sec:results_optical_thickness}), the single scattering albedo 
of the ring particles (Sect.~\ref{sec:results_albedo}), 
and the outer ring radius (Sect.~\ref{sec:results_ring_size}. 
To put the results into context we also performed a simple case 
study of the possibly ringed planet HIP~41378~f (Sect.~\ref{sec:case_study}). 

Our computations show a number of features indicative of a ring. 
In general, a ring (that is not seen edge-on), leaves two (different) 
peaks in the total 
flux of the system flux curve, with one due to the light that is reflected
by the ring and the other due to the diffusely transmitted light.
This was also reported by \citet{Arnold2004} and \citet{dyudina2005phase}. 
Including polarization, however, allows us to expand upon this and show that the increase in flux due to the ring that caused the second peak has a substantially lower degree of polarization. The lower degree of polarization of the light reflected or transmitted by the ring comes from the difference between molecules and particles and is therefore almost always present. This dichotomy between higher flux but a lower degree of polarization is therefore a key signature of a ring and was also reported in a recent study by \cite{Lietzow2023}. A notable exception to this would be a planet that shows large, seasonal activity, as was presented by \cite{dyudina2005phase} (without polarization). What sets the effect of rings apart from such a planet are the large changes in flux and degree of polarization around the ring-plane crossings, when the ring is illuminated edge-on. Therefore these sharp changes in the flux and degree of polarization curves would be something to look for in future observations. Lastly, due to the occultation of the ring in front of the planet or the shadow cast onto the planet, the received flux can be lower than expected, even when the ring is seen edge-on.

While each of the system parameters that were varied has distinct effects on the light curve we demonstrated that there is also a big overlap between them. Especially when it comes to the properties of the ring such as optical thickness, particle albedo, and size. Some of these properties, like the size of the ring and its optical thickness, also have an effect on the stability of the ring. So in order to draw a good conclusion, future observations should perform thorough fits of the parameter space to extract as much information as possible. Such a fit should include eccentric orbits, which is already possible using {\Pryn}, and different planetary atmospheres as was done by \cite{Lietzow2023}, which is not yet possible. The variation of the latter is interesting to combine with observations in different wavelength bands, in particular at wavelengths where gases in the planetary atmosphere absorb, and where the planet is thus dark, which could highlight the presence of rings and help to characterize them. 

The case study showed that it is currently impossible to directly observe such a system, it might be with the next generation of space telescopes like LUVOIR \citep{LUVOIR2019LUVOIR}, HabEx \citep{gaudi2018}, or the recently by NASA announced Habitable Worlds Observatory (HabWorlds or HWO). We show that it would indeed be beneficial for those missions to use polarimetry as it allows for the identification of properties like the optical thickness and particle albedo of exorings, something not possible using only flux measurements.

We conclude that a ring leaves unique features in the reflected 
and polarized fluxes of a planet as it orbits its star. 
In order to identify rings, observations 
along the planetary orbit would be needed, in particular aiming to 
cover the ring-plane crossings, where sharp changes in the system's 
signals should be detectable. 
Combining photometry and polarimetry would help to characterize the
physical properties of rings, such as their size, orientation, and 
optical thickness, which would allow us to better understand planet 
formation and the formation and evolution of the rings around the 
planets in our Solar System.

%% file: Xappendix.tex
\appendix

\section{Calculating the angles associated with the spangles}\label{appendix:1}
%%%%%%%%%%%%%%%%%%%%%%%%%%%%%%%%%%%%%%%%%%%%%%%%%%%%%%%%%%%%%%%%%%%%%%%%%%%%%%%%%
% Figure 2: Side view of the planet and ring
%%%%%%%%%%%%%%%%%%%%%%%%%%%%%%%%%%%%%%%%%%%%%%%%%%%%%%%%%%%%%%%%%
\begin{figure}
\centering
\includegraphics[width=0.49\textwidth]{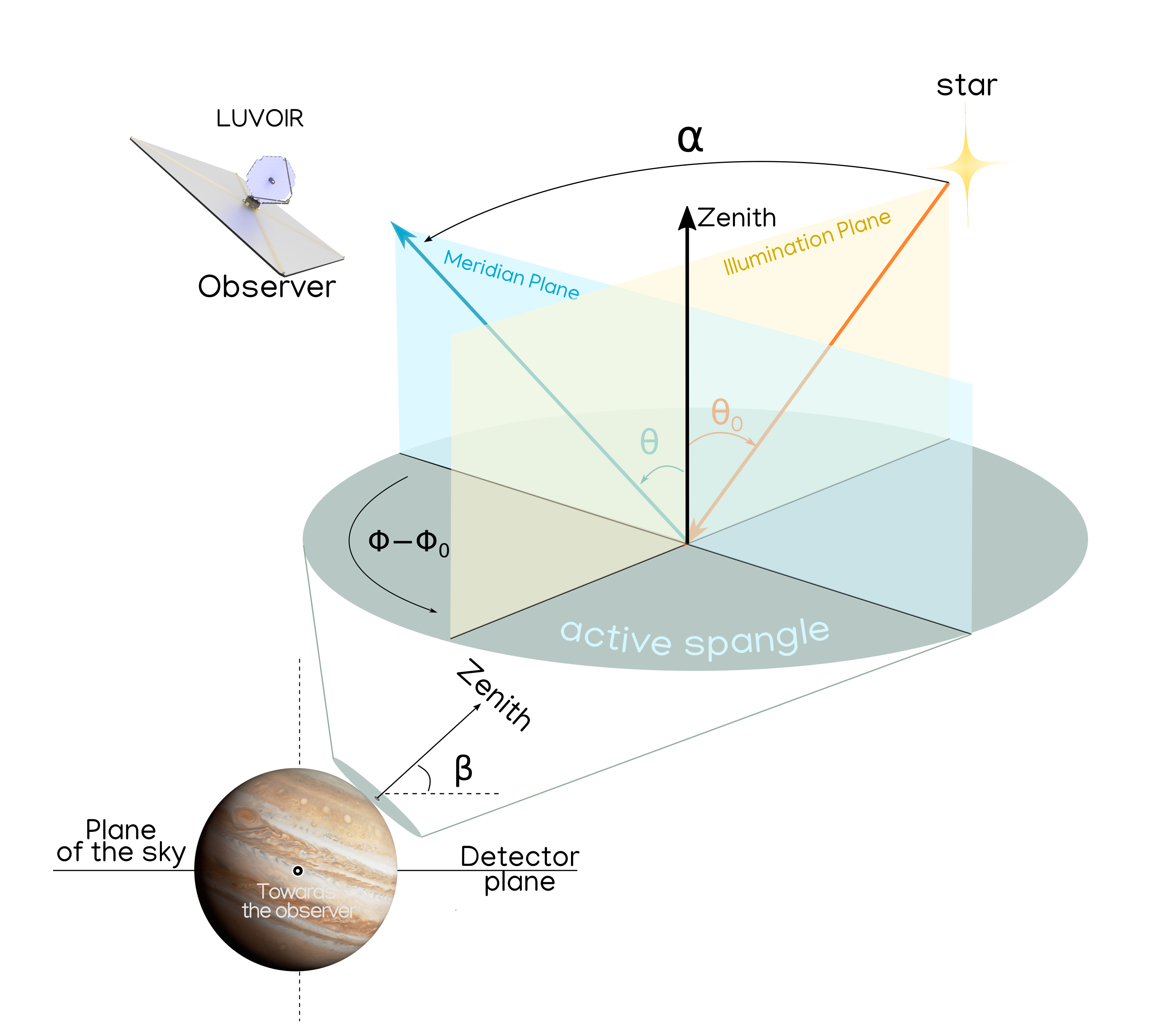}
\caption{Spangle geometry: orientation and angles involved in the calculation of stellar light scattering by an individual spangle, along with the observer's orientation (here, a LUVOIR-type telescope is the observer). }
\label{fig:spangle_geometry}%
\end{figure}
%%%%%%%%%%%%%%%%%%%%%%%%%%%%%%%%%%%%%%%%%%%%%%%%%%%%%%%%%%%%%%%%%
%%%%%%%%%%%%%%%%%%%%%%%%%%%%%%%%%%%%%%%%%%%%%%%%%%%%%%%%%%%%%%%%%%%%%%%%%%%%%%%%%

In Fig.~\ref{fig:spangle_geometry} we show the geometry of light scattering for a single spangle. For each spangle, we have to compute the local stellar zenith angle $\theta_0$,
defined as the angle between the local zenith direction and the direction towards
the star; the local viewing zenith angle $\theta$, which is the angle between
the local zenith direction and the direction towards the observer; and 
the local azimuthal difference angle $\phi - \phi_0$, taken as the angle between
the plane that contains the direction towards the local zenith and the 
direction towards the observer, and the plane that contains the direction 
towards the local zenith and the direction of propagation of the incident
starlight. Angle $\phi - \phi_0$ is measured rotating in the clockwise 
direction when looking towards the local zenith \citep[see][]{haan1987adding}. 
For a description of our computation of $\phi-\phi_0$, see Appendix~\ref{appendix:2}. 

In all cases, we assume that the incident direction of the starlight is the same across the planet and the ring (the local incident angles depend of course on the spangle). This assumption would not hold for computations
for planets with very wide rings and/or that are very close ($\ll0.1$~AU) to their star.

Given angles $i$, $\gamma$, and $\lambda_{\rm r}$ (which are part of the user-provided input to the package), {\Pryn}
computes, for each $\nu$ along the planetary orbit, the phase 
angle $\alpha$, and for each spangle on the planet and the ring, the angles 
$\theta_0$, $\theta$, and $\phi - \phi_0$. Note that unlike the planet spangles, 
all ring spangles have the same illumination and viewing geometries at a
given value of $\nu$. Then, taking into account the radius of the planet, the inner and
outer radii of the ring, and the direction to the star, {\Pryn} computes the state of the spangle, namely, whether it is illuminated by the star and visible to the observer; whether
it is in the shadow of the ring (for planet-spangles) or the planet (for 
ring-spangles); and/or whether it is occulted by the ring (for planet-spangles)
or the planet (for ring-spangles) (for a discussion on the key concept of ``spangle state'' in {\Pryn}, see the original paper by \citealt{zuluaga2022bright}).

%%%%%%%%%%%%%%%%%%%%%%%%%%%%%%%%%%%%%%%%%%%%%%%%%%%%%%
%%%%%%%%%%%%%%%%%%%%%%%%%%%%%%%%%%%%%%%%%%%%%%%%%%%%%%
\section{Calculating the ring-plane crossings}\label{appendix:3}
The location of the ring-plane crossings can be calculated by solving
\begin{equation}
\label{eq:ring_plane_crossing}
    {\bf n}_{\rm r} \cdot {\bf n}_{\rm star}(\nu_{\rm rp}) = 0\ ,
\end{equation}
with ${\bf n}_{\rm r}$ the normal vector to the ring, $\nu_{\rm rp}$ the true anomaly of the ring-plane crossings. 
and ${\bf n}_{\rm star}$ the normal vector pointing from the planet to the star.
We may express these locations in terms of the key angles:
\begin{align*}
    \sin{(\lambda_{\rm r})}& \cos{(\gamma)} \sin{(\nu_{\rm rp})}\ + \\ &\cos{(\nu_{\rm rp})}\ \Bigl[ \cos{(\lambda_{\rm r})} \cos{(\gamma)} \sin{(i)} - \sin{(\gamma)} \cos{(i)} \Bigl] = 0.
\end{align*}
In Fig.~\ref{fig:diff_positions}, for instance, where
$\lambda_{\rm r}=0^\circ$, the crossings occur at $\nu=90^\circ$ and 
270$^\circ$ (the rings are colored white there).

%%%%%%%%%%%%%%%%%%%%%%%%%%%%%%%%%%%%%%%%%%%%%%%%%%%%%%
%%%%%%%%%%%%%%%%%%%%%%%%%%%%%%%%%%%%%%%%%%%%%%%%%%%%%%
\section{Comparison with a Lambertian reflecting planet}\label{appendix:4}
\begin{figure}
    \centering
    \includegraphics[width=\hsize]{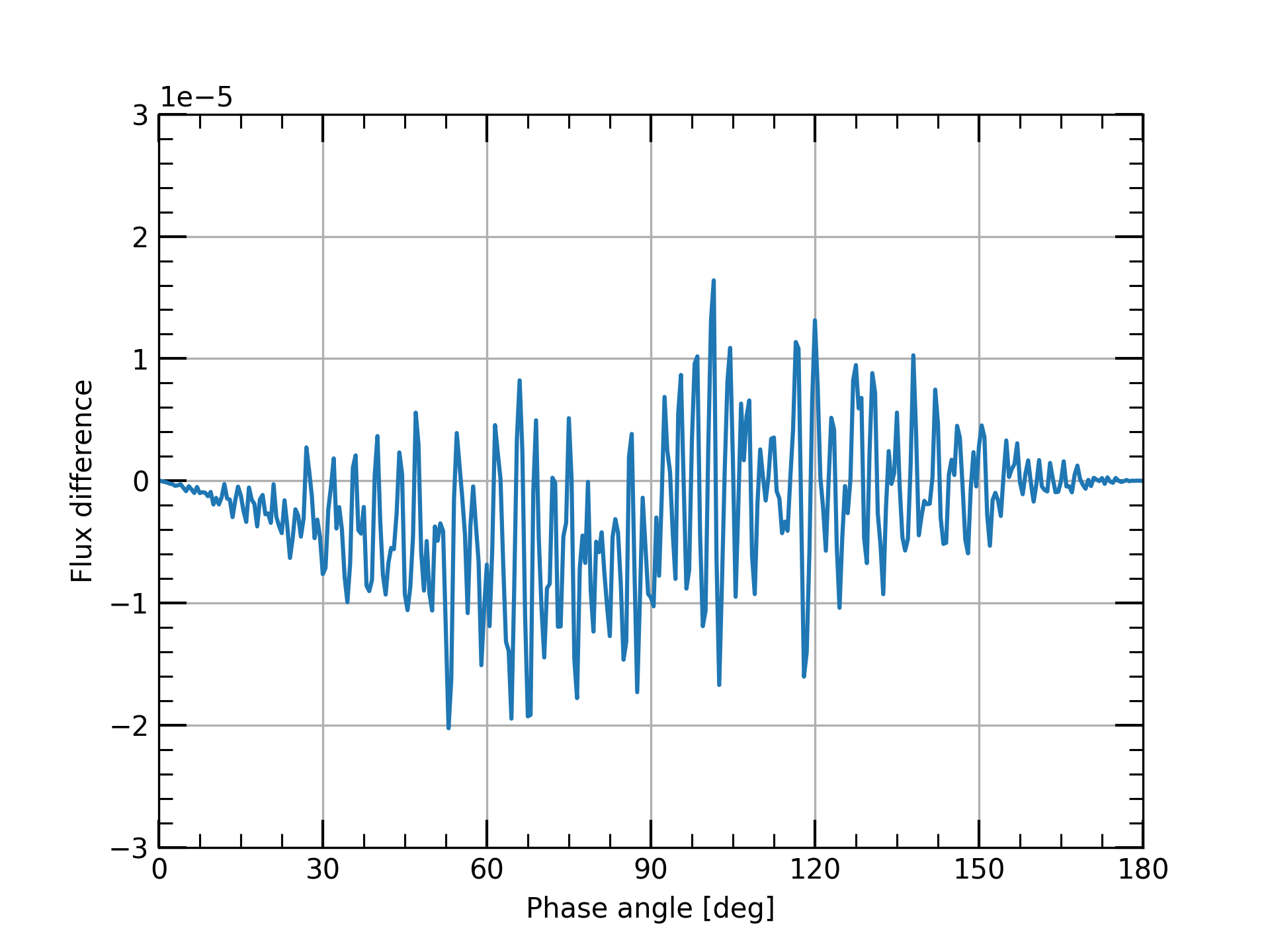}
    \caption{Difference in flux between theory and model for a Lambertian planet with 10,000 spangles.}
    \label{compare:diff_model}
\end{figure}

To validate the correct implementation of disk integration and geometry the total reflected flux is compared against the analytical expression for the phase function of a Lambertian reflecting planet with a surface albedo of $A_s$,
\begin{equation}
    F(\alpha) = \frac{2A_s}{3\pi}\left(\sin{\alpha} + (\pi - \alpha)\cos{\alpha} \right) .
\end{equation}
In figure \ref{compare:diff_model} this theoretical model, with a surface albedo of 1.0, is compared to a Lambertian reflecting planet modeled by {\Pryn}. The differences are very small and also show no systematic errors.

%%%%%%%%%%%%%%%%%%%%%%%%%%%%%%%%%%%%%%%%%%%%%%%%%%%%%%
%%%%%%%%%%%%%%%%%%%%%%%%%%%%%%%%%%%%%%%%%%%%%%%%%%%%%%

%%%%%%%%%%%%%%%%%%%%%%%%%%%%%%%%%%%%%%%%%%%%%%%%%%%%%%
%%%%%%%%%%%%%%%%%%%%%%%%%%%%%%%%%%%%%%%%%%%%%%%%%%%%%%
\section{Computing $\phi-\phi_0$ and $\beta$}\label{appendix:2}

%%%%%%%%%%%%%%%%%%%%%%%%%%%%%%%%%%%%%%%%%%%%%%%%%%%%%%
\begin{figure}
\centering
    \includegraphics[width=\hsize]{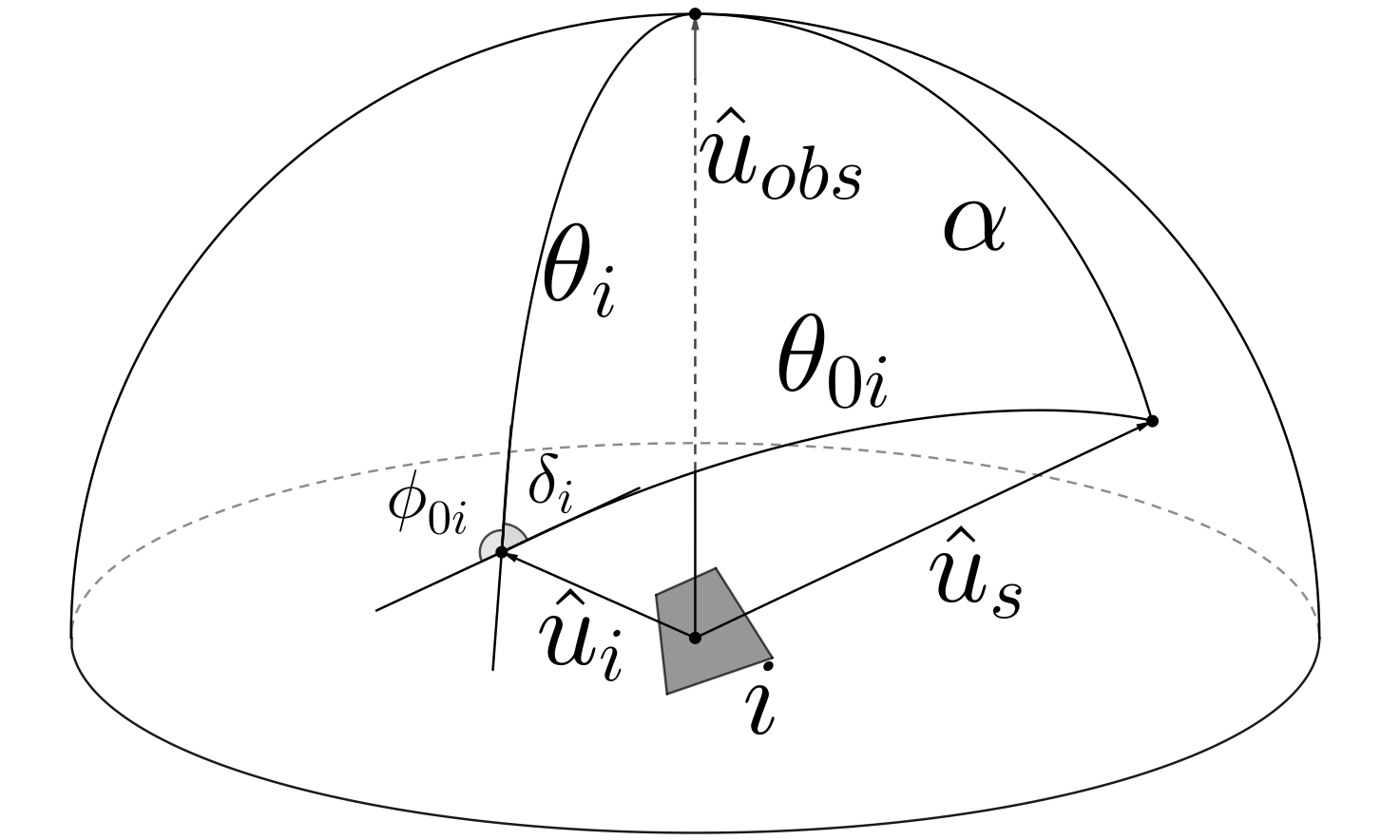}
    \caption{The angles and unit vectors associated with each individual spangle $i$. They are: the phase angle $\alpha$, direction to observer $\Hat{u}_{\rm obs}$, direction to the star $\Hat{u}_{\rm s}$, normal vector $\Hat{u}_{i}$, illumination angle $\theta_{i,0}$, viewing angle $\theta_i$, azimuthal angle $\phi_{0,i}$, and intermediate angle $\delta_i$. The rotation angle $\beta_i$ is not shown here.}
    \label{fig:zenith_angles}
\end{figure}
%%%%%%%%%%%%%%%%%%%%%%%%%%%%%%%%%%%%%%%%%%%%%%%%%%%%%%
We describe here the calculations for $(\phi - \phi_0)$ and $\beta$ in more detail. Starting with the azimuthal difference angle, and followed by the reference plane rotation angle $\beta$.

The azimuthal angles ($\phi$ and $\phi_0$) can be defined with respect to any plane containing the local z-axis which makes the plane that also contains the observer an obvious choice since this eliminates one of the two angles that need to be calculated, namely $\phi$. Thus an expression is needed for $\phi_0$ which then automatically also becomes an expression for $\phi - \phi_0$. Using the spherical law of cosine an expression can be found for an intermediate angle we name $\delta$ which is used to find $\phi_{0}$ with respect to the plane containing the direction to the observer and local z-axis,
\begin{align}
   \phi_{i,0} &= \pi - \delta_i \ , \\
   \delta_i   &= \arccos{\left( \frac{\cos{\alpha} - \cos{\theta_{i,0}} \cos{\theta_i} }{\sin{\theta_{i,0}} \sin{\theta_i} } \right)}\ ,
\end{align}
where the subscript $i$ stands for the $i^{\text{th}}$ spangle.

Care has to be taken, however, in making sure the angle that is found using the above equation is for a rotation that is clockwise when looking in the positive local zenith direction. Whether the found angle needs to be adjusted depends generally on the orientation of the spangle, be it planetary or ring, and the direction of the star-light. The dependency on the location of the star can be understood from Fig.~\ref{fig:zenith_angles} where $\alpha$ will change during the orbit, eventually moving to the other side of the plane formed by $\Hat{u}_i$ and $\Hat{u}_{obs}$, the normal vector of the spangle and the normal vector pointing to the observer respectively. For planetary spangles, the angles are modified as
\begin{align}
    (\phi_i - \phi_{i,0}) &= \phi_{i,0} \ , \\
    (\phi_i - \phi_{i,0}) &= -\phi_{i,0} \ , \quad \text{if}\quad y_i^{scat} < 0 \ , 
\end{align}
with $y_i^{scat}$ the $y$-location of the spangle with respect to the planetary scattering plane. Where the criterion there is to compensate for the flipping of the normal vector for spangles on the "southern" hemisphere. The dependency on the location of the star is already incorporated into the $y$-location with respect to the planetary scattering plane.

\begin{figure}
% \centering
    \includegraphics[width=9cm]{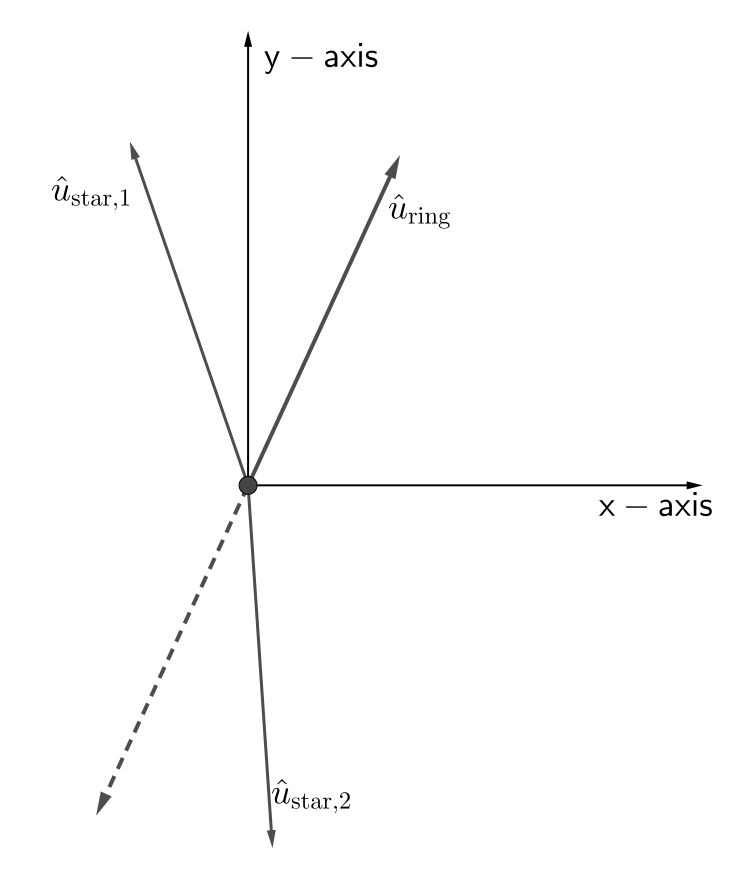}
    \caption{Illustration of the condition for changing the calculation of $\beta$, described in the text.}
    \label{fig:phi_ring_cond}
\end{figure}

Since all ring spangles have the same normal vector there is no dependency on their location on the ring. There is, however, a dependency on the orientation of the ring since it determines the locations in the orbit where the flip in the rotation direction happens. The locations of these flips are defined as the place where the plane containing the star, the planet, and the observer and the plane containing the normal vector of the ring and the observer are parallel. A 2D projection of the problem is shown in Fig.~\ref{fig:phi_ring_cond} with the two vectors in the direction of the star~$\Hat{u}_{\text{star, (1,2)}}$ representing two moments in the orbit. 

Two situations arise, depending on the orientation of the ring. If the normal vector of the ring~$\Hat{u}_{\text{ring}}$ is pointing in the $^+y$ direction,
\begin{equation}
    (\phi_i - \phi_{i,0}) = -\phi_{i,0} \ , 
\end{equation}
when the vector pointing to the star~$\Hat{u}_{\text{star}}$ is on the $^- x$ side of the plane formed by $\Hat{u}_{\text{ring}}$ and the observer. If $\Hat{u}_{\text{ring}}$ is pointing in the $^-y$ direction,
\begin{equation}
    (\phi_i - \phi_{i,0}) = -\phi_{i,0} \ , 
\end{equation}
when $\Hat{u}_{\text{star}}$ is on the $^+ x$ side of the plane formed by that same plane.

%%%%%%%%%%%%%%%%%%%%%%%%%%%%%%%%%%%%%%%%%%%%%%%%%%%%%%
%%%%%%%%%%%%%%%%%%%%%%%%%%%%%%%%%%%%%%%%%%%%%%%%%%%%%%

\noindent The angle between the local meridian plane and the detector plane, $\beta$, is also calculated differently for ring spangles compared to planetary spangles. For planetary spangles, the angle is only dependent on their location on the planet. With the detector plane as the reference plane, $\beta$ is calculated as
\begin{align}
   x_i y_i \geq 0:\ &\beta_i = \arctan{\frac{y_i}{x_i}}, \\
   x_i y_i < 0:\ &\beta_i = \pi + \arctan{\frac{y_i}{x_i}},
\end{align}
where the coordinates $(x_i,y_i)$ are in the observer reference frame. The addition of $\pi$ is there to make sure the angle that is calculated rotates in the correct direction.

Because all ring spangles have the same normal vector the rotation angle does not depend on the location of the individual spangles. Using the spherical cosine law we can find an equation that instead depends on the orientation of the ring
\begin{align}
   \beta &= \arccos{\left( \frac{\cos{\sigma}}{\sin{\theta_i}} \right)} ,\\
   \beta &= \pi - \arccos{\left( \frac{\cos{\sigma}}{\sin{\theta_i}} \right)} , \quad \text{if}\quad \Hat{u}_{ring}^y < 0 \ ,
\end{align}
with $\sigma = \text{arctan2}\left( \Hat{u}_{ring}^z / \Hat{u}_{ring}^x \right) $ the angle the normal vector makes with the $x$-axis and $\Hat{u}_{ring}^z$ the $z$-component of the normal vector. Again the condition of $\Hat{u}^{ring}_y < 0$ is added to make sure the rotation direction stays the same.